\documentclass[usenatbib]{mnras}

\usepackage{newtxtext,newtxmath}

\usepackage[T1]{fontenc}

\DeclareRobustCommand{\VAN}[3]{#2}
\let\VANthebibliography\thebibliography
\def\thebibliography{\DeclareRobustCommand{\VAN}[3]{##3}\VANthebibliography}

\usepackage{graphicx}	
\usepackage{amsmath}	
\usepackage{CJKutf8}
\usepackage{pifont}
\usepackage{enumerate}
\usepackage[dvipsnames]{xcolor}
\usepackage{float}
\usepackage{multirow}
\usepackage{caption}
\usepackage{orcidlink}

\newcommand{\RomanNumeral}[1]
{\MakeUppercase{\romannumeral #1}}

\newcommand{\angstrom}{\text{\normalfont\AA}}

\newcommand{\cmark}{\ding{51}}%
\newcommand{\xmark}{\ding{55}}%

\title[WASP-12~b atmosphere]{PEPSI's non-detection of escaping hydrogen and metal lines adds to the enigma of WASP-12~b}

\author[A. Pai Asnodkar et al.]
{\parbox{\textwidth}{Anusha Pai Asnodkar\orcidlink{0000-0002-8823-8237}$^{1}$ \thanks{E-mail: \texttt{paiasnodkar.1@osu.edu}} 
Ji Wang (王吉)\orcidlink{0000-0002-4361-8885}$^{1}$,
Madelyn Broome$^{2}$,
Chenliang Huang (黄辰亮)\orcidlink{0000-0001-9446-6853}$^{3}$,
Marshall C. Johnson\orcidlink{0000-0002-5099-8185}$^{1}$,
Ilya Ilyin$^{4}$,
Klaus G. Strassmeier\orcidlink{0000-0002-6192-6494}$^{4}$,
Adam Jensen$^{5}$}
\\ \\ \\
\parbox{\textwidth}{
$^{1}$The Ohio State University, McPherson Laboratory, 140 W 18th Ave., Columbus, OH 43210, USA \\
$^{2}$Department of Astronomy and Astrophysics, University of California Santa Cruz, 1156 High Street, Santa Cruz, CA 95064, USA \\
$^{3}$Shanghai Astronomical Observatory, Chinese Academy of Sciences, Shanghai 200030, People’s Republic of China \\
$^{4}$Leibniz-Institute for Astrophysics Potsdam (AIP), An der Sternwarte 16, D–14482 Potsdam, Germany \\
$^{5}$Department of Physics and Astronomy, University of Nebraska at Kearney, 2502 19th Avenue, Kearney, NE 68849, USA \\
}}

\date{Accepted XXX. Received YYY; in original form ZZZ}

\pubyear{2024}

\begin{document}
\begin{CJK*}{UTF8}{gbsn}
\label{firstpage}
\pagerange{\pageref{firstpage}--\pageref{lastpage}}
\maketitle 

\begin{abstract}
WASP-12~b is an ultra-hot Jupiter (UHJ) of special interest for atmospheric studies since it is on an inspiraling orbit in an extreme environment of intense radiation and circumstellar gas. Previously claimed detections of active mass loss from this planet are controversial across the literature. To address this controversy, we obtain two new transit observations of WASP-12~b with the optical high-resolution PEPSI spectrograph on the Large Binocular Telescope. Contrary to previous work, we do not observe planetary H$\alpha$ absorption and rule out the amplitude of previously reported detections. Our non-detection may be limited by the sensitivity of our data or could indicate weaker mass loss than suggested by previous studies. We conduct injection-recovery experiments to place constraints on the radial extent of WASP-12~b's escaping atmosphere as probed by Balmer lines, but find that our data do not have the sensitivity to probe down to the planet's Roche Lobe. Using physically motivated models of atmospheric escape, we explore upper limit constraints on the planet's mass-loss rate and deem the data quality in the wavelength regime of Balmer lines insufficient to determine a physically meaningful constraint. We also conduct a spectral survey of other optical absorbers to trace atmospheric circulation but detect no additional absorption. We conclude that previous claims of H$\alpha$ absorption from the atmosphere of WASP-12~b should be reevaluated. Given the anticipated line strength of Balmer/optical features, observing the atmosphere of this faint target will require stacking more observations even with the largest telescope facilities available.

\end{abstract}




\section{Introduction}

Among the diverse menagerie of exoplanets discovered to-date, WASP-12~b \citep{Hebb2009} is one of the most extreme worlds ($R_\mathrm{p} = 1.9\ R_{\mathrm{jup}}$, $M_\mathrm{p} = 1.47\ M_{\mathrm{jup}}$, $P \approx 1.09$ days, $T_{\mathrm{eq}} = 2580$ K; \citealt{Collins2017}), displaying exotic phenomena unlike anything in our own Solar System. It was the first system to confirm predictions from theory that some close-in giants may have decaying orbits due to tidal interactions with their host stars \citep{Maciejewski2016, Yee2020, Turner2021}. With an inspiral timescale of $\leq2.9 \pm 0.14$ Myr \citep{Efroimsky2022}, WASP-12~b is an ideal testbed for probing the tidal quality and, in effect, the interiors of irradiated gas giants \citep{Barker2009, Patra2017, Weinberg2017, Millholland2018, Maciejewski2018, Bailey2019, Maciejewski2020}. Recently, \cite{Efroimsky2022} demonstrated that WASP-12~b's orbital decay is consistent with tidal dissipation and its tidal quality factor is comparable to Jupiter's. Such gravitational influences distort WASP-12~b into a prolate, disrupted planet being devoured by its host star as indicated by the imprint of the planet's geometry on its  light curve \citep{Li2010, Akinsanmi2024}. 

Spectroscopic investigation of the WASP-12 system generally corroborates the aforementioned findings. The host star's lack of the line core emission from tracers of chromospheric activity (Mg \RomanNumeral{2} h\&k and Ca \RomanNumeral{2} H\&K) that is typical for stars of comparable spectral type (between late-F to early-G) and age ($\sim2$ Gyr) has been suggested as evidence of additional absorption from a circumstellar torus of material \citep{Bonomo2017, Haswell2012, Haswell2018, Debrecht2018}. This is further supported by WASP-12 b's early ingress near-ultraviolet absorption reported in \citet{Fossati2010}, suggesting the disk material is perhaps stripped from the planetary atmosphere, although the possibility of ejected plasma debris from outgassing of a Trojan satellite or exomoon has also been proposed \citep{Haswell2012, Fossati2013, Debrecht2018, Kislyakova2016, Ben-Jaffel2014}. A deeper investigation of the planet's atmosphere in the ultraviolet (UV) during transit provides indirect evidence for the presence of numerous metals (Al \RomanNumeral{2}, Fe \RomanNumeral{2}, Mg \RomanNumeral{2}, Mn \RomanNumeral{1}/\RomanNumeral{2}, Na \RomanNumeral{1}, Sc \RomanNumeral{2}, Sn \RomanNumeral{1}, V \RomanNumeral{2}, and Yb \RomanNumeral{2}) in its exosphere, indicative of a hydrodynamic outflow \citep{Fossati2010, Haswell2012}. 

Transmission spectroscopy in the optical at medium resolution (R$\sim$15,000 with the High Resolution Spectrograph on the 10-m Hobby-Eberly Telescope) revealed H$\alpha$ and Na \RomanNumeral{1} absorption from the planetary atmosphere, but no H$\beta$ or Ca \RomanNumeral{1} \citep{Jensen2018}. Their work remains inconclusive on the radial extent of the $n=2$ hydrogen population probed by their H$\alpha$ detection and whether or not the probed atmosphere overfills the planetary Roche lobe, an independent indication of active hydrodynamic escape. More recent work by \citet{Czesla2024} reports no extended atmospheric H$\alpha$ absorption from transit observations of WASP-12~b with the CARMENES high-resolution spectrograph (R$\sim$94,600) on the 3.5-m telescope at the Calar Alto Observatory. \citet{Kreidberg2018} reports a notable non-detection of helium, another tracer of atmospheric escape, in the planet's exosphere from the metastable triplet feature in the infrared; this non-detection is further supported by \citet{Czesla2024}.

In this work, we extend the optical exploration of WASP-12 b's atmosphere to higher resolution (R $\sim$ 130,000) and larger telescope aperture (2 $\times$ 8.4 m, with an effective aperture size of 11.8 m) with the PEPSI spectrograph on the Large Binocular Telescope (LBT). Higher resolution can provide clarity on velocity information that is lost at lower resolutions to constrain the dynamics of a planetary outflow as well as circulation in the upper atmosphere. Furthermore, the LBT's larger telescope aperture relative to the Hobby-Eberly Telescope and the CARMENES instrument is more sensitive to low signal-to-noise ratio (SNR) planetary signals. We conduct transmission spectroscopy with PEPSI-LBT to follow-up the investigation of H$\alpha$ in \cite{Jensen2018} as well as other metal species at optical wavelengths. In \S \ref{sec:observations}, we describe the details of the two transit observations we obtained with PEPSI. In \S \ref{sec:methods}, we outline our pipeline for reducing the data. In \S \ref{sec:hydrogenExtent}, we discuss constraints on the hydrogen envelope's radial extent from injection-recovery tests. Similarly, we attempt to place constraints on the planet's mass-loss rate in \S \ref{sec:hydrogenMassLoss}. \S \ref{sec:otherSpecies} describes our search for other optical absorption in WASP-12 b's atmosphere and we discuss our constraints in the context of previous work and other comparable systems. We present our conclusions in \S \ref{sec:conclusions}.

\section{Observations}
\label{sec:observations}

\begin{table*}
\caption{Observing log of WASP-12~b transit observations: columns provide the night number, date of observation, telescope observing mode (default is binocular, but can be monocular in unexpected circumstances), UTC start and end times of observing, number of exposures per night, exposure times in seconds for PEPSI blue and red arms, and average single exposure SNR per pixel for PEPSI blue ($\lambda$4800 -- 5441 \angstrom) and red arms ($\lambda$6278 -- 7419 \angstrom).}
\centering
\begin{tabular}{lllllllllll}
\hline
\hline
Night & Date & Observing Mode & $t_{\mathrm{start}}$ (UTC) & $t_{\mathrm{end}}$ (UTC) & $N_{\mathrm{obs}}$ & $t_{\mathrm{exp, blue}}$ (s) & $t_{\mathrm{exp, red}}$ (s) & Airmass & $\overline{\mathrm{SNR_{blue}}}$ & $\overline{\mathrm{SNR_{red}}}$\\ 
\hline
1 & 2020-11-22 & Binocular & 07:12:38.0 & 13:15:34.6 & 24 & 900 & 900 & 1.0013 - 1.451 & 97.3 & 135.5 \\
2 & 2020-12-27 & Monocular & 04:47:32.2 & 10:18:53.8 & 22 & 900 & 900 & 1.0015 - 1.2761 & 67.1 & 94.2 \\
\hline
\end{tabular}
\label{tab:datasets}
\end{table*}

We conduct high-resolution transmission spectroscopy of WASP-12~b using the optical Potsdam Echelle Polarimetric and Spectroscopic Instrument \citep[PEPSI;][]{Strassmeier2015} on the Large Binocular Telescope \citep[LBT; two 8.4-m mirrors, effective aperture of 11.8 m in binocular mode;][]{Wagner2008}. We use data from cross-dispersers \RomanNumeral{3} ($\sim$4800--5441 \text{\AA}, R=130,000) and \RomanNumeral{5} ($\sim$6278--7419 \text{\AA}, R=130,000) for the necessary wavelength coverage to observe Balmer line features (H$\alpha$ and H$\beta$). From the standard PEPSI pipeline \citep{Ilyin2000, Strassmeier2018}, we obtain continuum-normalized, order-stitched 1D spectra for every exposure, each corrected for solar barycentric motion. These spectra from the pipeline are ready to be directly adopted by the procedure described in \S \ref{sec:methods}. We analyze two nights of observations in this work; see Table \ref{tab:datasets} for details about the observations.

Both data sets include observations taken during transit as well as several immediately before and after the transit to establish an out-of-transit baseline. We adopt an exposure time of 900 s for both PEPSI arms. On Night 1, the continuum SNR per exposure ranged between 77 to 113 in the blue arm and 106 to 156 in the red arm. On Night 2, the continuum SNR ranged between 52 to 77 in the blue arm and 73 to 107 in the red arm. As indicated in Table \ref{tab:datasets}, Night 2 observations have lower signal-to-noise because the LBT was configured in monocular mode for simultaneous observation with the LUCI \citep{Seifert2003} instrument (these data are not relevant for this work). We converted all observation timings from the provided $\mathrm{JD_{UTC}}$ timings to $\mathrm{BJD_{TDB}}$ using the Time Utilities\footnote{\url{https://astroutils.astronomy.osu.edu/time/utc2bjd.html}} online software tool \citep{Eastman2012} to make them comparable with the ephemeris of the WASP-12 system given in \cite{Wong2022}.

\section{Methods}
\label{sec:methods}

We follow a procedure for transmission spectrum construction similar to that in \citet{PaiAsnodkar2022}. To extract transmission spectra of WASP-12~b's atmosphere, we: 
\begin{enumerate}
\item perform least-squares deconvolution (LSD) to recover stellar radial velocities (RVs),
\item perform a Keplerian RV curve fit to recover the orbital properties of the host star,
\item shift all observations to the stellar rest-frame,
\item divide all observations by a combined stellar spectrum constructed from out-of-transit observations,
\item apply the SYSREM \citep{Tamuz2005} algorithm to remove systematic effects.
\end{enumerate}
We will proceed to describe each of these steps in further detail.

We are interested in the probing the dynamics of the planet's atmosphere. This will require spectroscopically measuring the radial velocity of the planet's atmospheric absorption signature relative to the orbital motion of the planet. The first step towards achieving this requires shifting to the stellar rest-frame to remove the stellar component, which requires knowing the systemic velocity and RV semi-amplitude of the host star. We extract the orbital properties of the host star from the out-of-transit observations taken on both nights of observation using LSD and RV fitting as described in \S 3.1 of \citet{PaiAsnodkar2022}. First, we generate template spectra of the star in the IDL software \texttt{Spectroscopy Made Easy (SME)} \citep{Valenti1996, Valenti2012} at 21 different limb-darkening angles using the \textit{VALD3} linelist for a $T_{\mathrm{eff}}$ = 6360 K star as an input. We integrate across the stellar disk, which we treat as a grid of $0.01 R_\star \times 0.01 R_\star$ cells, and continuum-normalize to construct a stellar template spectrum. We then conduct LSD \citep{Kochukhov2010, Donati1997} to recover the empirical broadening profiles of the out-of-transit observations of the host star. The broadening kernel includes a Gaussian component from instrumental broadening \citep{Strassmeier2018} and a rotational broadening component as defined analytically in \citealt{gray_2005}. We globally fit the empirical profiles with a model that is the convolution of the Gaussian and rotational components. The centroids of the model fit to each empirical profile is the radial velocity of the star at the orbital phase corresponding to the observation. The centroids are determined according to a circular orbital solution:
\begin{equation}
v_\star(t) = K_\star \sin \Big( \frac{t - t_0}{p} \Big) + v_{\mathrm{sys}}
\label{eq:RV_star}
\end{equation}
The free parameters in our model of the line broadening profiles are the stellar linear limb-darkening coefficient, $v \sin{i_\star}$, stellar RV semi-amplitude ($K_{\mathrm{\star}}$) and the systemic velocity measured by the PEPSI instrument ($v_{\mathrm{sys}}$). A multiplicative scaling factor and an additive offset are included as nuisance parameters in the fitting. The multiplicative scaling factor rescales the analytical kernel to match the amplitude of the empirical deconvolved line profiles. The additive offset is necessary because the empirical deconvolved line profiles may not have a baseline centered at 0 due to a lack of flux conservation between the template and observed spectra. $K_{\mathrm{\star}}$ and $v_{\mathrm{sys}}$ are our parameters of interest for shifting our observations to the stellar rest-frame. In our fitting procedure, we restrict the linear limb-darkening coefficient between 0.3867 and 0.4903 according to the range of plausible values given in \citealt{Claret2017}. Note that linear limb darkening laws are known to be an oversimplification of stellar intensity profiles. However, this simplified broadening profile is sufficient to determine the centroids of the observed line profiles to constrain $K_\star$ within $\lesssim$0.05 km s$^{-1}$ uncertainties as described in the proceeding bootstrapping procedure. This uncertainty on $K_\star$, which we use to shift the spectra to the stellar rest-frame, is well within the $\sim 1$ - 2 km s$^{-1}$ uncertainties on the radial velocity of the planet's atmospheric absorption we expect with PEPSI \citep{PaiAsnodkar2022}.

\par We apply least-squares fitting to determine the best-fit parameters. To estimate parameter uncertainties, we bootstrap the residuals of the flat region of the deconvolved kernel, add the samples to the best-fit model kernel, and refit the line profiles. We obtain a stellar RV semi-amplitude of $K_\star = 0.3288^{+0.0158}_{-0.0392}$~km~s$^{-1}$ and a systemic velocity of $v_{\mathrm{sys}} = 19.275^{+0.0243}_{-0.0066}$~km~s$^{-1}$. This is consistent with the $20.62 \pm 1.44$~km~s$^{-1}$ reported on the Gaia archive \cite{Gaia2021}. We caution that our value for $K_\star$ may be skewed since we are limited by the phase coverage of our observations; \citet{Collins2017} reports $0.2264 \pm 0.0041$ km s$^{-1}$, which differs from our measurement by 6.3$\sigma$. However, this discrepancy in $K_\star$ should not significantly affect the shift to the stellar rest-frame over the limited phase coverage of our observations during transit, at least not beyond the measurement uncertainties of the planet's atmospheric dynamics which we can measure within 1 - 2 km s$^{-1}$ at best as previously mentioned.

Next, we shift all of our spectra for a given night of observation to the stellar rest-frame according to our empirically derived orbital parameters of the star. We then take the error-weighted mean of the out-of-transit observations to obtain a combined stellar spectrum. We divide the combined stellar spectrum out of all of our observations to remove the dominant stellar component and reveal the absorption features from the planet's atmosphere in transmission. This operation should yield roughly flat spectra for the out-of-transit observations and potentially spectra with absorption features from the planetary atmosphere for the in-transit phases. Unlike in \citet{PaiAsnodkar2022}, we do not see traces of a Doppler shadow from the Rossiter-McLaughlin effect \citep{Rossiter1924, McLaughlin1924} in the resulting transmission spectra since the host star WASP-12 is a slow rotator with $v\sin{i_\star} = 1.6 ^{+0.8}_{-0.4}$~km~s$^{-1}$ \citep{Albrecht2012}. This is consistent with \citet{Husnoo2011}; therefore, we do not model the Doppler shadow in subsequent analysis.

\begin{figure*}
\begin{minipage}{0.48\linewidth}
    \centering
    \includegraphics[width=\textwidth]{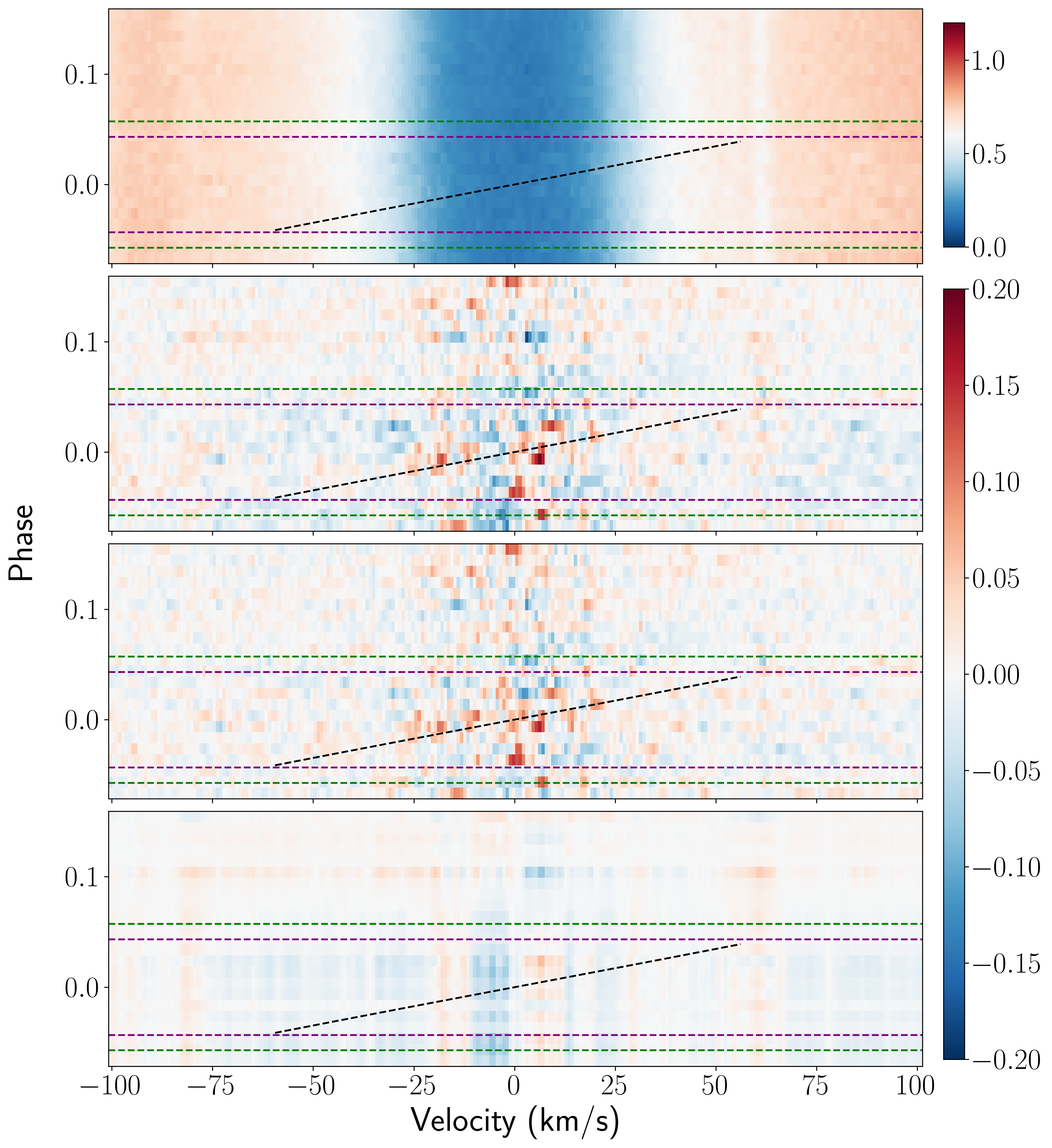} \\
    \label{fig:Halpha_SYSREM}
    (a)
\end{minipage}\hfill
\begin{minipage}{0.48\linewidth}
    \centering
    \includegraphics[width=\textwidth]{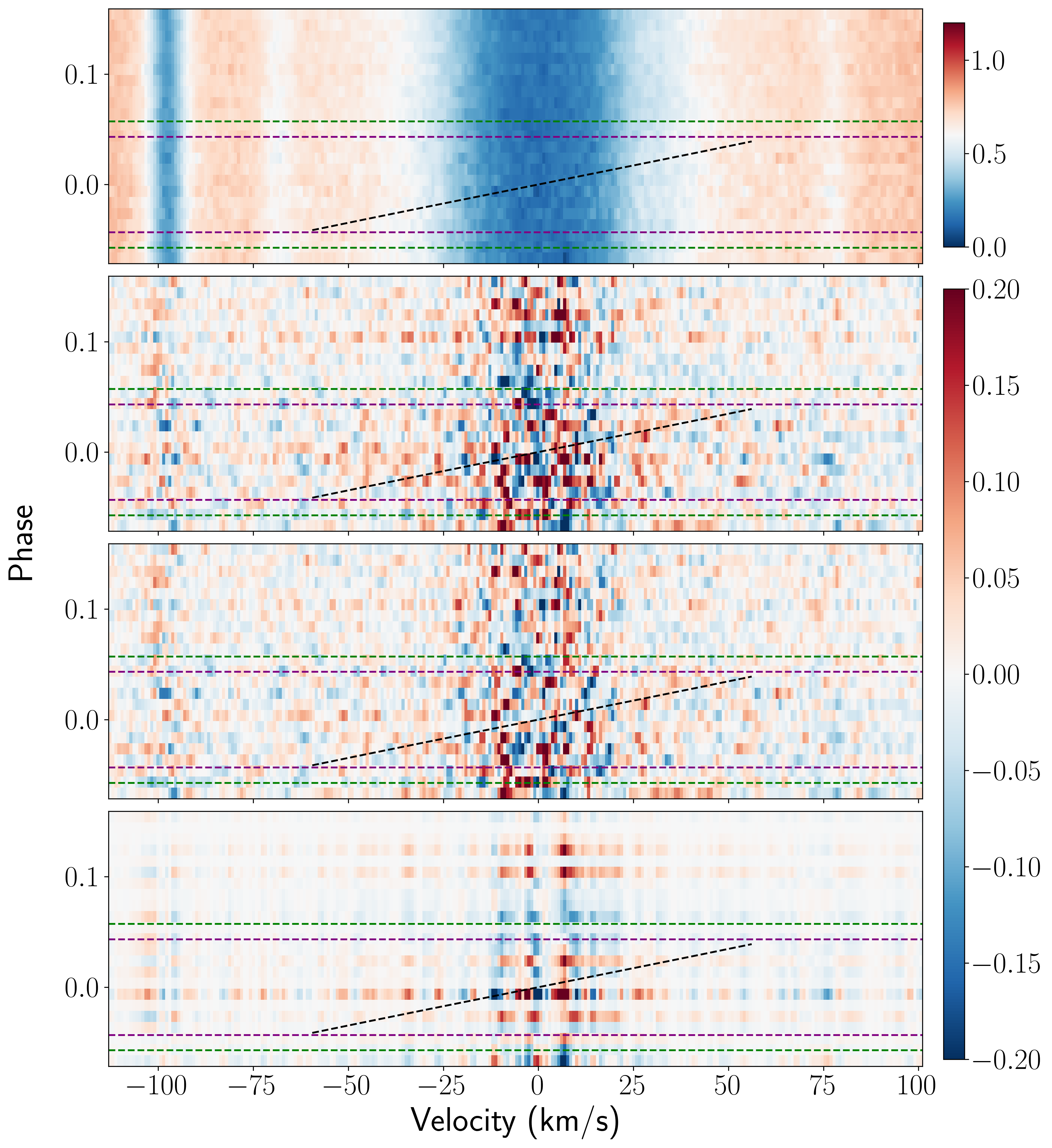}\\
    \label{fig:Hbeta_SYSREM}
    (b)
\end{minipage}
\caption{(a) Non-detection of planetary atmospheric absorption around H\text{$\alpha$} line (6562.83 $\angstrom$) in Night 1 data. The top panel are the order-stitched, continuum-normalized stellar spectra from the PEPSI pipeline where the colorbar represents flux after continuum normalization. The second panel are the spectra with the stellar component removed, i.e. the out-of-transit combined stellar spectrum has been divided out and the result is subtracted by 1 such that the baseline flux of the star fluctuates around zero; red (values greater than zero) indicates excess emission while blue (values less than zero) indicates excess absorption. The third panel shows the spectra from the second panel after applying SYSREM (2 systematics, 100 iterations per systematic). The fourth and bottom panel displays the difference between the second and third panels. In all panels, the green horizontal dashed lines indicate the phases of 1st and 4th contact of transit while the purple horizontal dashed lines indicate the phases of 2nd and 3rd contact. The black dashed line maps out the planet's expected orbital radial velocity over the course of its transit based on the system parameters provided in \citet{Collins2017}. The assignment of orbital phases to each observation is dependent on our adopted mid-transit timing from \citet{Wong2022}, which has an uncertainty of 6.22 seconds. If the mid-transit time were off by 1$\sigma$, it would shift the track horizontally by $\sim$0.1 km s$^{-1}$. (b) Same as (a), but around H\text{$\beta$} line (4861.34 $\angstrom$).}
\label{fig:Balmer_data}
\end{figure*}

In addition to the lack of the Doppler shadow due to slow rotation, another deviation from the data reduction in \citet{PaiAsnodkar2022} is the addition of the SYSREM algorithm to this pipeline. The host star is not a rapid rotator, so we can see deep absorption lines from the stellar photosphere in our observations. Since the scatter in stellar spectral flux across observations is roughly uniform over the wavelengths of observation, the signal-to-noise at the core of a stellar absorption feature is significantly lower than in the continuum. As a result, when we divide out the combined stellar spectrum, we obtain noisy residual streaks across observations at the wavelengths corresponding to a stellar absorption line (see Figure \ref{fig:Balmer_data}). Similar streaks have been observed previously in PEPSI transmission observations of the 55 Cnc system \citep{Keles2022} and in CARMENES observations of our target system WASP-12 \citep{Czesla2024}. 

We attempt to mitigate this undesirable artifact by employing the SYSREM algorithm  \citep{Tamuz2005} to filter out systematics across observations. SYSREM operates like an extension of principle component analysis to iteratively identify linear systematic effects. The performance of the algorithm is determined by two parameters: 1) the number of systematics being identified and 2) the number of chi-squared minimization iterations for each systematic. As shown by the vertical streaks in Figure \ref{fig:Balmer_data}, this procedure is insufficient to completely remove the artifact. However, it is more effective at removing systematics across cross-correlated data (see \S \ref{sec:other_species}). Furthermore, the streak is relatively stationary in velocity space around 0 km s$^{-1}$ (since the observations have been shifted to the stellar rest-frame and in general the star should be close to stationary in the context of the velocities spanned in Figure \ref{fig:Balmer_data}) while the planet's absorption signature should range roughly between -60 to 60~km~s$^{-1}$ over the course of its transit across the stellar disk (assuming no additional radial velocity shifts from atmospheric dynamics). Thus we expect these two signatures to be distinct in velocity space and we should still be able to see traces of the planet's atmospheric absorption if any such features are sufficiently high signal. For example, Figure 3 of \citet{Mounzer2022} shows an example of excess sodium absorption from KELT-11 b's atmosphere during transit in spite of an obstructive stellar line core artifact.

SYSREM also minimizes telluric contamination because the spectra are shifted by at most $\sim$0.42 km s$^{-1}$ relative to each other to convert from Earth's rest-frame to WASP-12's stellar-frame, so telluric absorption is relatively stationary across observations. Furthermore, tellurics are more prevalent in the PEPSI red arm observations, which we only use to search for H$\alpha$ absorption, a feature in a wavelength window relatively free of strong tellurics \citep{Smette2015}. We only use blue arm observations (in which tellurics are negligible) for the cross-correlation analysis in \S \ref{sec:other_species}. Thus, tellurics are not a significant concern in this work.

\section{Constraining the radial extent of WASP-12~\lowercase{b}'s hydrogen envelope}
\label{sec:hydrogenExtent}

\subsection{Balmer line absorption}
Since ultra-hot Jupiters like WASP-12~b experience stellar irradiation that is strong enough to thermally dissociate molecular hydrogen, neutral hydrogen is expected to be a dominant species in their upper atmospheres. Hydrogen Balmer lines in planetary atmospheres can probe neutral hydrogen at high altitudes and the escape regime \citep{Yan2018, Cauley2019, Casasayas-Barris2019, Jensen2018, Cabot2020, Yan2021}. In our wavelength regime with PEPSI cross-dispersers \RomanNumeral{3} and \RomanNumeral{5}, we can observe the H$\alpha$ (6562.83 $\angstrom$) and H$\beta$ (4861.34 $\angstrom$) features \citep{Wiese2009}. Just by eye, we do not observe any excess absorption attributable to planetary absorption around either of these wavelengths (e.g. see Figure \ref{fig:Balmer_data}); likewise, we do not observe planetary Balmer line absorption in publicly available archival HARPS-North transit data sets (2017-12-23 and 2018-11-14)\footnote{\url{http://archives.ia2.inaf.it/tng/}}. We attempt a quantitative recovery of Balmer line signals in transmission using Bayesian parameter estimation and a model of the planet's atmospheric absorption that is described further in \S \ref{sec:Balmer_retrieval}, but still recover no signal. This is consistent with the absence of atmospheric H$\alpha$ absorption from CARMENES data reported in \citet{Czesla2024}, but in tension with the strong detection in \citet{Jensen2018}. See \ref{sec:Balmer_retrieval} for a quantitative framework for placing upper limit constraints on the radial extent of WASP-12~b's hydrogen envelope.

We note that \cite{Kreidberg2018} and \citet{Czesla2024} report non-detections of helium, another commonly adopted tracer of atmospheric escape. It has been noted that the metastable helium triplet can be insufficiently populated for atmospheres receiving low extreme-UV and X-ray irradiation from their host stars \citep{Sanchez-Lopez2022}. However, this is unlikely to be applicable to WASP-12~b considering the helium triplet has been observed in HD 209458~b \citep{Alonso-Floriano2019}, which orbits a G0 star of a similar spectral type as WASP-12~b's host. On the other hand, the UV flux of the host star WASP-12 may be anomalous for its spectral type as indicated by its lack of emission in the line cores of the  Mg \RomanNumeral{2} h\&k stellar activity tracers \citep{Fossati2011}, although this has also been speculatively attributed to absorption from a circumstellar disk of escaping gas from the planet or potential Trojan satellites. The metastable state of helium can also be depopulated in the most extreme UHJ atmospheres (like WASP-12~b) exposed to high near-UV flux \citep{Oklopcic2019}. 

\subsection{Injection-recovery analysis of Balmer lines}
\label{sec:Balmer_retrieval}
Although we do not observe any excess Balmer absorption from the planetary atmosphere during transit, we place upper limit constraints on the radial extent of WASP-12~b's hydrogen envelope based on the noise properties (scatter of the transmission spectra with the stellar component divided out) of our data sets. The aim of this is to determine if the planet's atmosphere is confined within its Roche lobe \citep{Eggleton1983} and is thus unlikely to be in the regime of strong hydrodynamic escape. We conduct an injection-recovery analysis, modelling planetary absorption observed with transmission spectroscopy as a Gaussian feature broadened by instrumental effects. Note that the true line profile from a hydrodynamic outflow will not be a pure Gaussian due to a combination of thermal, rotational, instrumental, and outflow expansion velocity broadening effects. However, we do not expect the precise line shape of the model to significantly affect our injection-recovery analysis given the data quality.

We first generate Gaussian signals to represent the planet's Balmer line absorption. Adopting the system parameters from \citet{Collins2017}, the translation from atmospheric radial extent $R_{\mathrm{ext}}$ to the depth of the Gaussian absorption feature $\delta$ is:
\begin{equation}
\delta = \frac{R_{\mathrm{ext}}^2 - R_{\mathrm{p}}^2}{R_\star^2}
\label{eq:absorption_depth}
\end{equation}

The widths of the Gaussian signals we inject are motivated by limits taken from observations and models of other UHJs. Across the literature, the observed full width at half-maximum (FWHM) of Balmer absorption from an UHJ ranges from $\sim20$--$50$ km s$^{-1}$ \citep{Casasayas-Barris2019, Wyttenbach2020, Borsa2021, Zhang2022}. We take 20 km s$^{-1}$ as the lower limit and adopt $60$ km s$^{-1}$ as the upper limit on the FWHM of the signals we inject according to a multispecies hydrodynamic model of a WASP-121~b  \citep{Huang2023}, a planet with properties very similar to WASP-12~b. The Gaussian signal is convolved in a flux-conserving manner (constant equivalent width) with an instrumental broadening kernel. PEPSI's instrumental broadening is captured by a Gaussian kernel with a full-width at half maximum in velocity units of $\frac{c}{R}$, where $c$ is the speed of light and $R$ is the spectral resolution of the data ($R\ =$ 130,000).
The resulting signal is subtracted from 1 (continuum) and injected (multiplied) into our normalized observations as an H$\alpha$ or H$\beta$ absorption feature, where the center of the signal is offset from the central wavelength according to the radial orbital motion of the planet; this offset calculation is analogous to Equation \ref{eq:RV_star}, except the planet's projected orbital velocity ($K_{\mathrm{pl}}$) is used in place of $K_\star$ and the planet's net radial velocity offset in the stellar rest-frame (set to 0 km s$^{-1}$) is used instead of $v_{\mathrm{sys}}$. In doing so, the noise properties of the observations are incorporated in the simulated data. We assume the planet's orbital motion is circular with projected orbital velocity $K_\mathrm{pl} = \frac{2 \pi a_{\mathrm{pl}}}{P} = 231.7$~km~s$^{-1}$ based on orbital parameters from \citet{Wong2022} and that there are no additional velocity offsets from atmospheric dynamics.

\begin{figure*}
\begin{minipage}{0.5\linewidth}
    \centering
    \includegraphics[height=\textwidth]{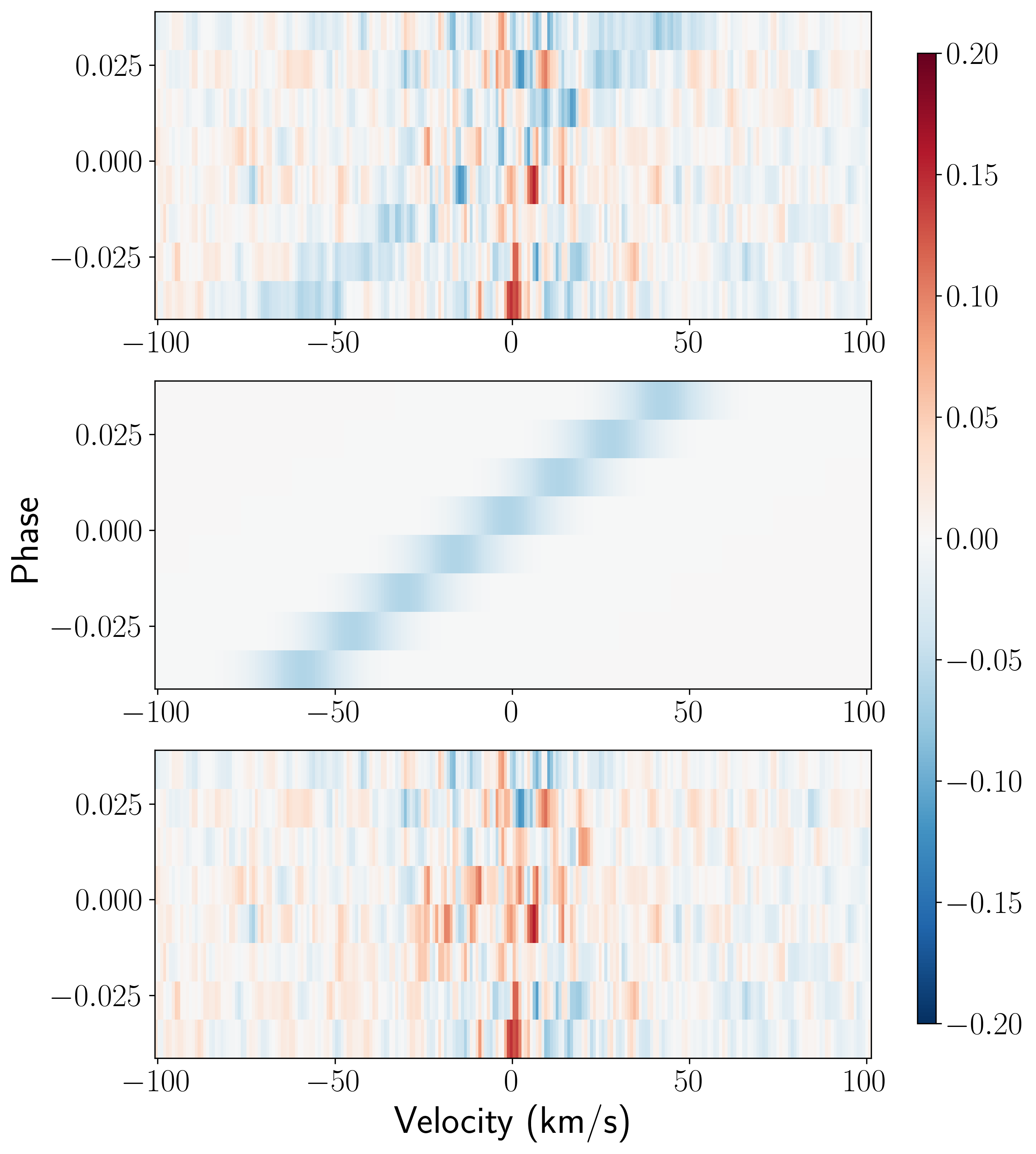} \\
    (a)
\end{minipage}\hfill
\begin{minipage}{0.5\linewidth}
    \centering
    \includegraphics[height=\textwidth]{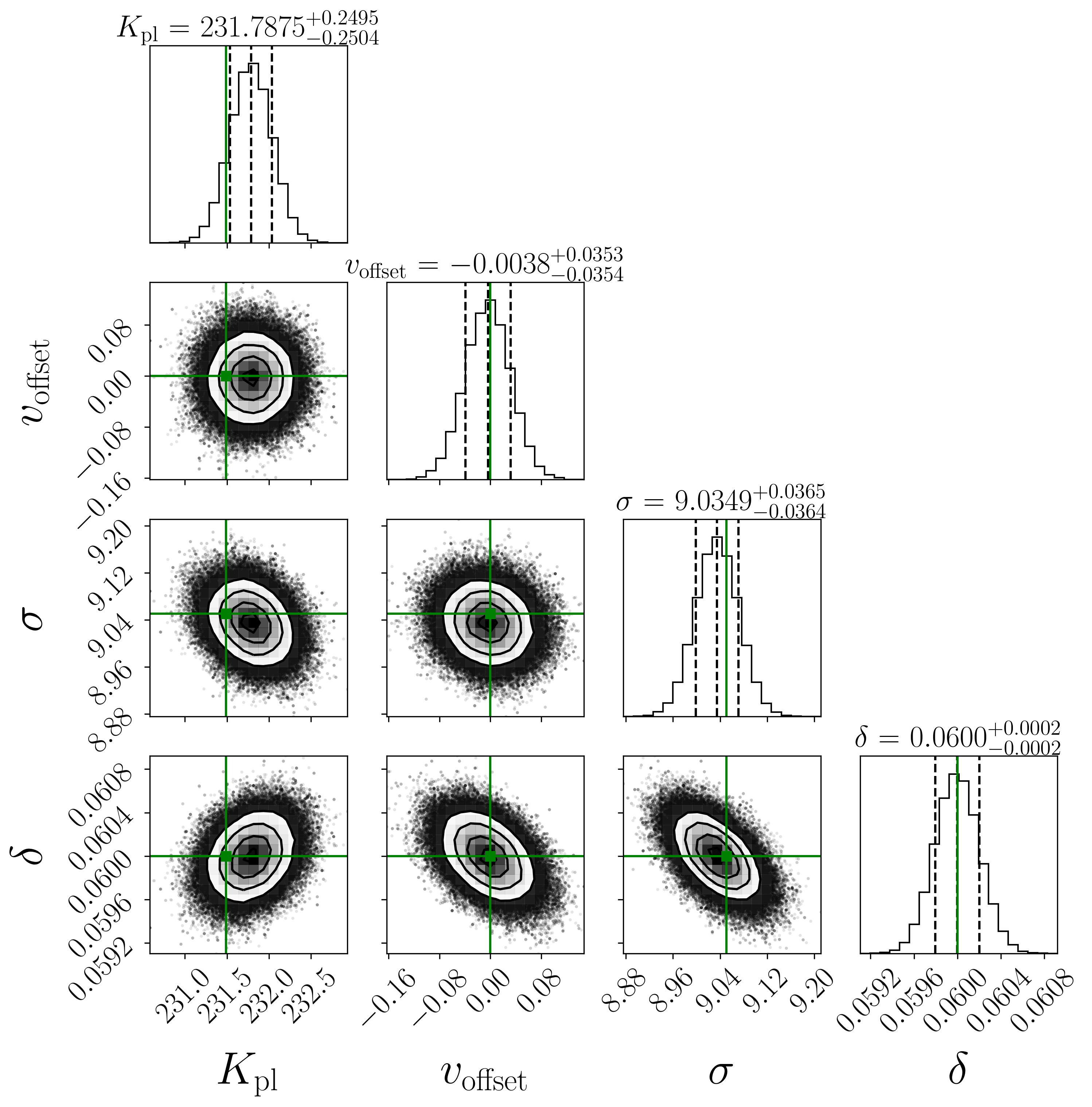}\\
    (b)
\end{minipage}
\caption{(a) Injection (top panel), recovery (model fit in middle panel), and residuals (difference map between injected data and model fit, bottom panel) of a planetary H$\alpha$ signal comparable to the detection in \citet{Jensen2018}. Note that this plot is showing a simulated H$\alpha$ feature incorporating the noise of our spectroscopic observations and does not depict a true, measured absorption feature in our PEPSI data. As in Figure \ref{fig:Balmer_data}, red indicates excess emission and blue indicates excess absorption. (b) Corner plot of parameter posterior distributions from MCMC sampling. The green lines identify the true injected parameters and fall within the $1\sigma$ regime of the sampled distributions, indicating that the injected signal is successfully recovered. The injected signal is retrieved with an SNR of 5$\sigma$.}
\label{fig:Jensen18Injection}
\end{figure*}

To find the upper limit constraint on the planet's radial extent, we use the root-finding routine (\texttt{scipy.optimize.root}) to minimize $R_{\mathrm{ext}}$ until the SNR of the injected feature in the simulated data is at least 3.  To calculate SNR, we first stack all in-transit simulated observations in the planet's rest-frame. We define SNR of the resulting stacked absorption feature as follows:
\begin{equation}
SNR = \frac{A}{RMS * FWHM}
\label{eq:SNR}
\end{equation}
In Equation \ref{eq:SNR}, the signal $A$ is the planetary absorption signal in the simulated data subtracted from the continuum and integrated over the velocity range that spans the FWHM of the signal. The denominator captures the noise, in which $RMS$ is the root mean square (RMS) error  of the residuals within the FWHM of the signal, i.e. 
\begin{equation}
RMS = \sqrt{\frac{\sum_{i = 1}^{N} (\mathrm{data}_i - \mathrm{model}_i)^2}{N}}
\label{eq:RMS}
\end{equation}
for N data points across the FWHM of the absorption line in all in-transit observations. To make the noise an area comparable to the integral we use for estimating the signal in Equation \ref{eq:SNR}, we scale the $RMS$ by $FWHM$, which is the full-width at half-maximum of the absorption line.  

To ensure the injected 3$\sigma$ signals can be retrieved amidst the observational noise, we perform model-fitting with MCMC sampling using \texttt{emcee} \citep{Foreman-Mackey2013}; see Figure \ref{fig:Jensen18Injection}. The model that we fit to the simulated data is the same model used to generate the injected signal. We adopt a Bayesian framework for sampling the parameter space with MCMC applying linearly uniform priors and marginalize over model parameters, i.e. $K_{\mathrm{pl}}$ (planet RV semi-amplitude), $v_{\mathrm{offset}}$ (net radial velocity offset of signal relative to the stellar rest-frame), $\sigma$ (Gaussian width), and $\delta$ (absorption depth) as shown in Figure \ref{fig:Jensen18Injection} as an example.  The ability to recover the injected model parameters is limited by the noise quality of the data. We use $-\frac{\chi^2}{2}$ as the log-likelihood of a given model and scale the priors by the number of elements in the observed flux map before summing the log-prior and log-likelihood to obtain the log-posterior probability. We run the \texttt{emcee} sampler with 10 walkers until the following convergence criteria are met: 1) the estimated autocorrelation time is 1\% of the chain length and 2) the estimated autocorrelation time has changed by less than 1\% , checking every 100 steps. We consider a signal successfully recovered if the posterior distributions of all parameters appear sufficiently Gaussian or converged.

Figure \ref{fig:Balmer_radialExtent} provides a summary of our upper limits on WASP-12~b's hydrogen envelope as probed by H$\alpha$ and H$\beta$ for both nights of observation.

\subsection{Upper limits on WASP-12~b's excited-state hydrogen radial extent are above the planet's Roche lobe}
\label{sec:radialExtent}

\begin{figure}
\centering
\includegraphics[width=0.5\textwidth]{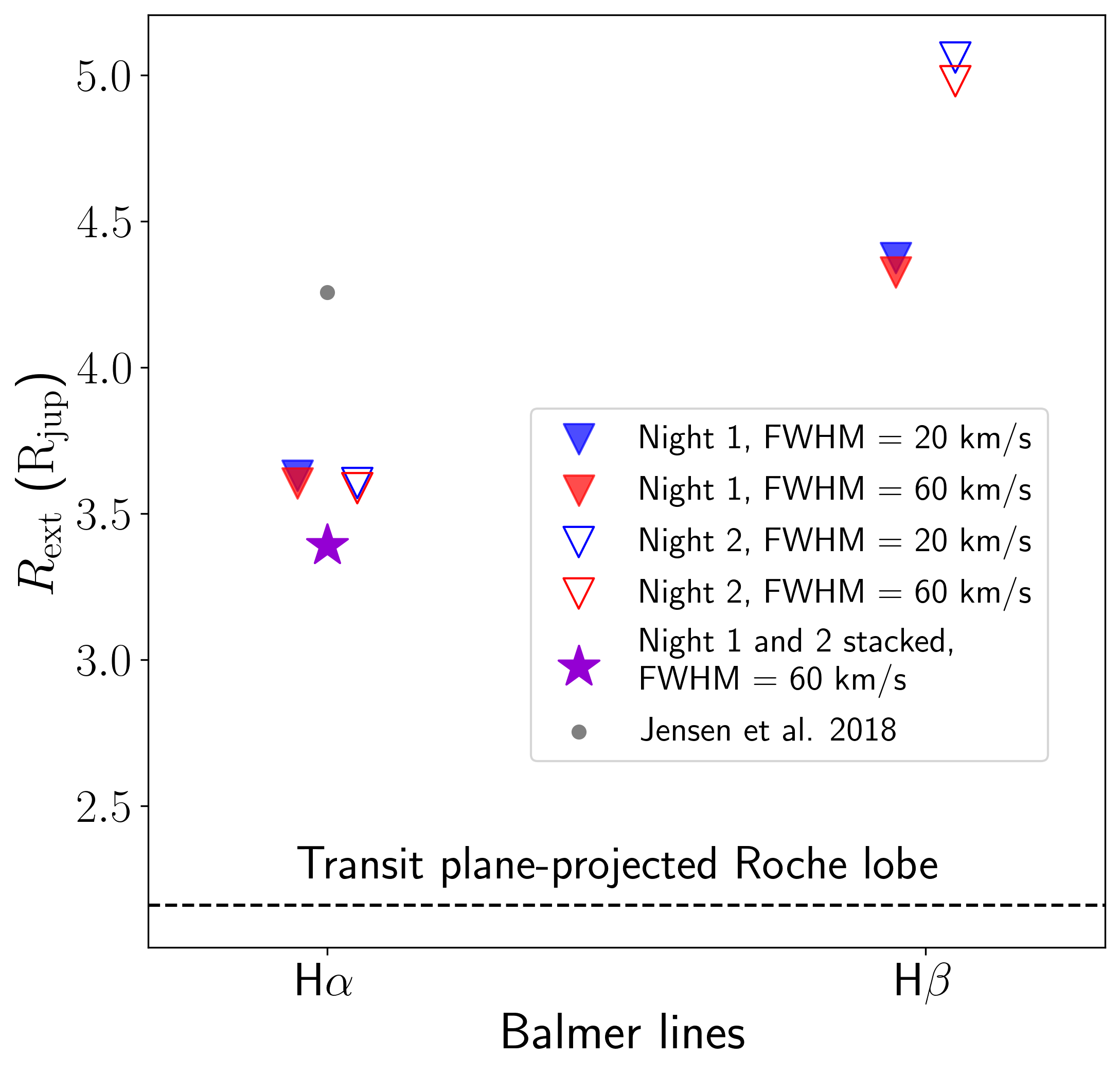}
\caption{Upper limits (3$\sigma$) on WASP-12~b's hydrogen envelope compared to: 1) the planet's Roche lobe projected onto the transit plane as perceived with transmission spectroscopy and 2) H$\alpha$ radial extent reported in \citet{Jensen2018}.}
\label{fig:Balmer_radialExtent}
\end{figure}

We define the lowest value of $R_{\mathrm{ext}}$ that yields a 3$\sigma$ signal as our empirical upper limit constraint on the radial extent of WASP-12~b's hydrogen envelope (see Figure \ref{fig:Balmer_radialExtent}). The disparity in the upper limit constraints on radial extent between different absorption features and data sets arises from the empirical noise corresponding to the wavelength regime of that feature for that night of observation. It is worth noting that the noise in the red arm observations near the wavelength regime of H$\alpha$ is very similar for both nights of observation, while the blue arm observations near H$\beta$ have larger scatter (by a factor of $\sim$1.5) in Night 2 observations than Night 1; this is why both nights yield similar upper limits on radial extent probed by H$\alpha$, but a smaller radial extent from H$\beta$ with Night 1 data than Night 2. Since the blue arm is generally noisier than the red arm, we get tighter constraints from the H$\alpha$ upper limits than H$\beta$.

From Figure \ref{fig:Balmer_radialExtent}, it is evident that for individual nights of observation, we are unable to constrain the radius of the planet's neutral hydrogen envelope down to its Roche lobe given our data quality. In this situation, the planet may have a hydrodynamically escaping atmosphere that extends beyond the Roche lobe but below our sensitivity, or it may have a weakly escaping atmosphere confined within the Roche lobe. As indicated by the purple star in Figure \ref{fig:Balmer_radialExtent}, even when we stack both nights, we find that at best the $\mathrm{FWHM} = 60$ km s$^{-1}$ scenario is observable in H$\alpha$ with $\geq$3$\sigma$ confidence if $R_{\mathrm{ext}} \geq 3.39\ R_{\mathrm{jup}}$, which is well above  WASP-12~b's transit plane-projected Roche Lobe of 2.16 $R_{\mathrm{jup}}$. 

The true conditions (temperature and hydrogen number density) in WASP-12~b's upper atmosphere remain elusive due to the absence of an observable absorption signature, so we are unable to conclusively make a statement about the status of WASP-12~b's atmospheric escape from our upper limit constraints on its Balmer line photosphere. In \S \ref{sec:hydrogenMassLoss}, we investigate the possibility that WASP-12~b's mass-loss may be much weaker than expected based on other UHJs.

\subsection{Tension with Jensen et al. 2018}
\label{sec:Jensen2018}

We inject an H$\alpha$ signal of a similar strength as the detection in \citet{Jensen2018} measured with the High-Resolution Spectrograph \citep[HRS;][]{Tull1998} on the Hobby-Eberly Telescope \citep[HET;][]{Ramsey1998, Hill2021}, in our Night 1 observations; see Figure \ref{fig:Jensen18Injection}. We set the absorption depth to 0.06 based on Figure 9 of \citet{Jensen2018}. We choose a Gaussian standard deviation corresponding to the planet's rotational velocity (assuming the planet is tidally synchronous with the star) at $R_{\mathrm{pl}}$ (9.05 km s$^{-1}$). This is narrower than any width we consider in our injection-recovery analysis in \S \ref{sec:radialExtent} and amounts to a width of $\sim$0.2 \angstrom; the equivalent width of the signal we inject is -29.8 m\angstrom, which is less than half the equivalent width of WASP-12~b's H$\alpha$ signature of -64.9 m\angstrom\ reported in \citet{Jensen2018}. Note that a real signal would appear narrower and deeper in PEPSI data than the HRS data from \citet{Jensen2018} because PEPSI (R $\sim$ 130,000) is higher resolution than HRS (R $\sim$ 15,000); however instrumental broadening preserves total flux, so the equivalent width of both signals should be the same. Since our injected signal's equivalent width is less than that of the H$\alpha$ feature presented in \citet{Jensen2018}, it should be more challenging to recover. Nevertheless, we are able to retrieve this signal with an SNR of 5$\sigma$. Therefore, we have the data quality to detect an H$\alpha$ absorption feature with the same properties as the strong signal detected in \citet{Jensen2018}, so our incongruent lack of such a detection challenges the collective understanding of the WASP-12 system across the literature.

\begin{figure}
\centering
\includegraphics[width=0.45\textwidth]{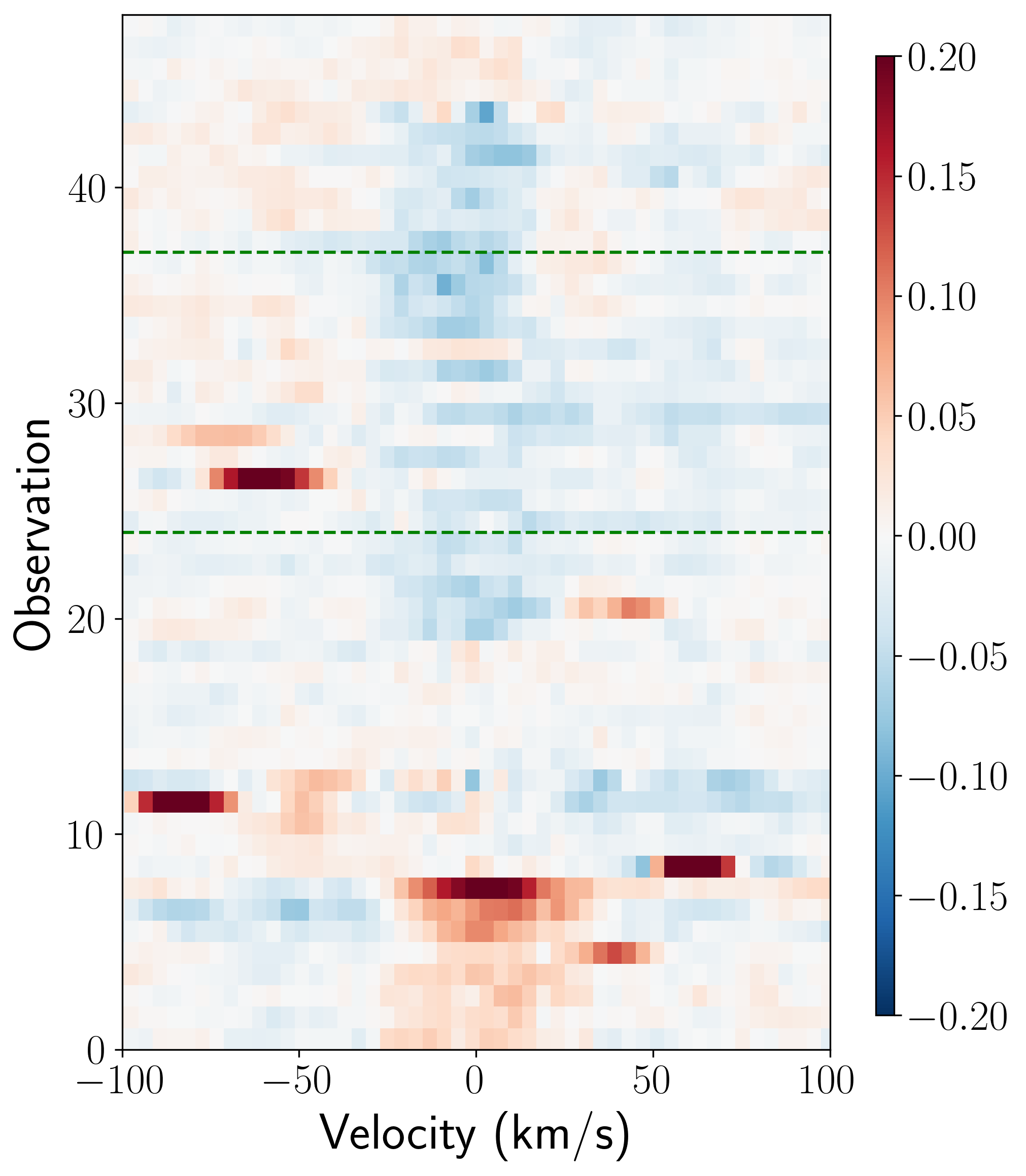}
\caption{Transmission spectra using the HET datasets from \citet{Jensen2018}, extracted by our pipeline. As in Figure \ref{fig:Balmer_data}, red indicates excess emission and blue indicates excess absorption. Horizontal green dashed lines correspond to 1st and 4th contact.}
\label{fig:Halpha_Jensen}
\end{figure}

To explore this discrepancy with \citet{Jensen2018} in further detail, we run their H$\alpha$ observations from HRS through our data analysis pipeline. Our extraction of the transmission spectrum can be seen in Figure \ref{fig:Halpha_Jensen}. Since these data are not uniformly sampled in phase, we present the spectra in order of increasing phase from bottom to top. The horizontal green dashed lines indicate 1st and 4th contact; thus observations between these lines were taken during transit. We correct for the 18.75 km s$^{-1}$  systemic radial velocity adopted in \citet{Jensen2018} according to \citet{Gaia2016}. These data present a potentially transit-correlated absorption signal as indicated by the blue streak. However, this signal is stationary in radial velocity, whereas a signal from a close-orbit planet like WASP-12 b should span a range of radial velocities between $\sim -60$ to 60 km s$^{-1}$ over the course of its transit due to the planet's orbital motion. Furthermore, the absorption depth is a factor of $\sim$4 to 12 times deeper than we expect from atmospheric escape models and what is observed from similar systems (see \S \ref{sec:hydrogenMassLoss}). While it may have planetary origins that can be explained by exotic physical mechanisms, we believe this signature is the residual from the removal of the noisy stellar line core, much like the artifact we see in our PEPSI transmission spectra for this system. This is not addressed in \citet{Jensen2018}. Thus while these data do appear to present potentially astrophysical signals, we believe the planetary origin of this signal remains to be determined and presents a challenge to theoretical models of atmospheric escape. 

\section{Upper limit constraint on hydrogen mass loss rate}
\label{sec:hydrogenMassLoss}

\subsection{Model injection-recovery with \texttt{p-winds}}
\label{sec:p-windsMethods}
The primary quantity of interest in studies of atmospheric escape is the mass loss rate of the planetary atmosphere ($\dot{M_{\mathrm{pl}}}$), which can shape the evolution of a planet over the course of its lifetime if sufficiently high. Constraining this mass-loss rate requires a physically-motivated model of the outflow structure and a radiative transfer scheme for estimating transmission spectroscopy signals. We adopt the open-source code \texttt{p-winds} \citep{DosSantos2022} for this purpose. This code assumes the outflow can be approximated by an isothermal, purely H-He Parker wind and solves for steady-state ionization balance. It also includes a module for ray-tracing and radiative transfer to calculate the in-transit spectrum. 

We note that this code is intended for modelling metastable He absorption \citep{Oklopcic2018} and does not include a routine for estimating Balmer series population levels. For this reason, we primarily use \texttt{p-winds} to estimate the outflow structure (velocities and individual species number densities as a function of radius) for a given mass loss rate. For this step, we: 1) use the planetary parameters from \citet{Collins2017}, 2) assume a hydrogen number fraction of 0.9, 3) assume solar abundances of carbon and oxygen (the two metal species included this model), and 4) scale the solar high-energy spectrum\footnote{\url{https://raw.githubusercontent.com/ladsantos/p-winds/main/data/solar_spectrum_scaled_lambda.dat}} for WASP-12~b's orbital configuration to estimate the wavelength-dependent photoionizing instellation upon the planet. We also enable \texttt{p-winds}'s functionality to include tidal effects since WASP-12~b's mass loss is expected to be dominated by Roche lobe overflow \citep{Koskinen2022}. Then we adopt the Monte Carlo-based framework in \citet{Huang2017} to estimate the $n=2$ state number densities from Equation 14 of \citet{Huang2017}. We provide the Balmer series number density structure as an input to the \texttt{p-winds} radiative transfer routine, along with the microphysical parameters (central wavelength, oscillator strength, and Einstein coefficient from \citet{Wiese2009}) of H$\alpha$ and H$\beta$ instead of the metastable helium triplet as originally intended. We generate models for two different mass-loss cases, $\dot{M_{\mathrm{pl}}} = 10^{10}$ g s$^{-1}$ (weak) and $\dot{M_{\mathrm{pl}}} = 10^{12}$ g s$^{-1}$ (moderate). The moderate mass loss case is defined based on observed mass loss rates of other UHJs, like KELT-9~b and KELT-20~b, which range between 10$^{12}$ and 10$^{13}$ g s$^{-1}$ \citep{Yan2018, Wyttenbach2020, Huang2023}. Furthermore, a hydrodynamic escape code we adopt (see \S \ref{sec:relaxed-ae}) predicts a mass-loss rate of $\sim 4 \times 10^{12}$ g s$^{-1}$ for this system.

After generating Balmer absorption signals with \texttt{p-winds}, we inject them in our observations to place upper limits on WASP-12~b's mass loss rate. As in \S \ref{sec:Balmer_retrieval}, we apply instrumental broadening to the signal before injection. We also perform rotational broadening since the 1D escape code does not incorporate the line-of-sight effects from planetary rotation. We follow \S4.3 of \citet{Huang2023} and build a rotational broadening kernel corresponding to the planetary rotational velocity at $R_{\mathrm{ext}}$ assuming WASP-12~b is tidally locked with its host star and obeys rigid body rotation. Since transmission spectroscopy only probes the transparent terminator region of the planet's atmosphere, we define the rotational broadening kernel $L_{\mathrm{rot}}(v)$ such that the value at a given line-of-sight projected rotational velocity of the planet $v$ is weighted by the cross-sectional length of the atmosphere illuminated at that velocity, i.e.:
\begin{equation}
L_{\mathrm{rot}} (v) =
\begin{cases}
    \sqrt{R_{\mathrm{ext}}^2\ \Big(1-\frac{v^2}{v_\mathrm{rot}^2}\Big)} - \sqrt{R_{\mathrm{core}}^2 - \frac{R_{\mathrm{ext}}^2 v^2}{v_\mathrm{rot}^2}} & \text{if } \frac{R_{\mathrm{ext}} v}{v_\mathrm{rot}} <  R_{\mathrm{core}} \\
    \sqrt{R_{\mathrm{ext}}^2\ \Big(1-\frac{v^2}{v_\mathrm{rot}^2}\Big)} & \text{if } \frac{R_{\mathrm{ext}} v}{v_\mathrm{rot}} \geq  R_{\mathrm{core}}
\end{cases}
\label{eq:rot_kernel}
\end{equation}
Here $v_\mathrm{rot}$ is the line-of-sight projected rotational velocity of the planet (we assume it is tidally-locked) at $R_{\mathrm{ext}}$. We adopt the approximation $R_{\mathrm{core}} \approx R_{\mathrm{pl}}$ because $R_{\mathrm{core}}$ is the radius at which the planet is opaque to all wavelengths. For $R_{\mathrm{ext}}$, we fit the absorption feature that comes out of \texttt{p-winds} (before applying velocity broadening from the outflow's expansion) with a Gaussian model to estimate $\delta$ and inverting Equation \ref{eq:absorption_depth} to solve for $R_{\mathrm{ext}}$, then apply Equation \ref{eq:rot_kernel}. We convolve the expansion velocity-broadened signal from \texttt{p-winds} with the rotational broadening kernel and instrumental kernel in a flux-conserving manner (constant equivalent width) for completeness, although we find neither of these broadening effects significantly impact the depth or shape of the signal. Then we inject the signal into our observations assuming it is shifted in velocity by the projected orbital motion of the planet exclusively. The injected data is fit with the model used in \S \ref{sec:hydrogenExtent}. The signal is considered observable if it is recovered with SNR > 3. 

\subsection{Upper limits on WASP-12~b's mass loss rate from \texttt{p-winds} are inconclusive}

\begin{figure}
    \centering
    \includegraphics[width=0.5\textwidth]{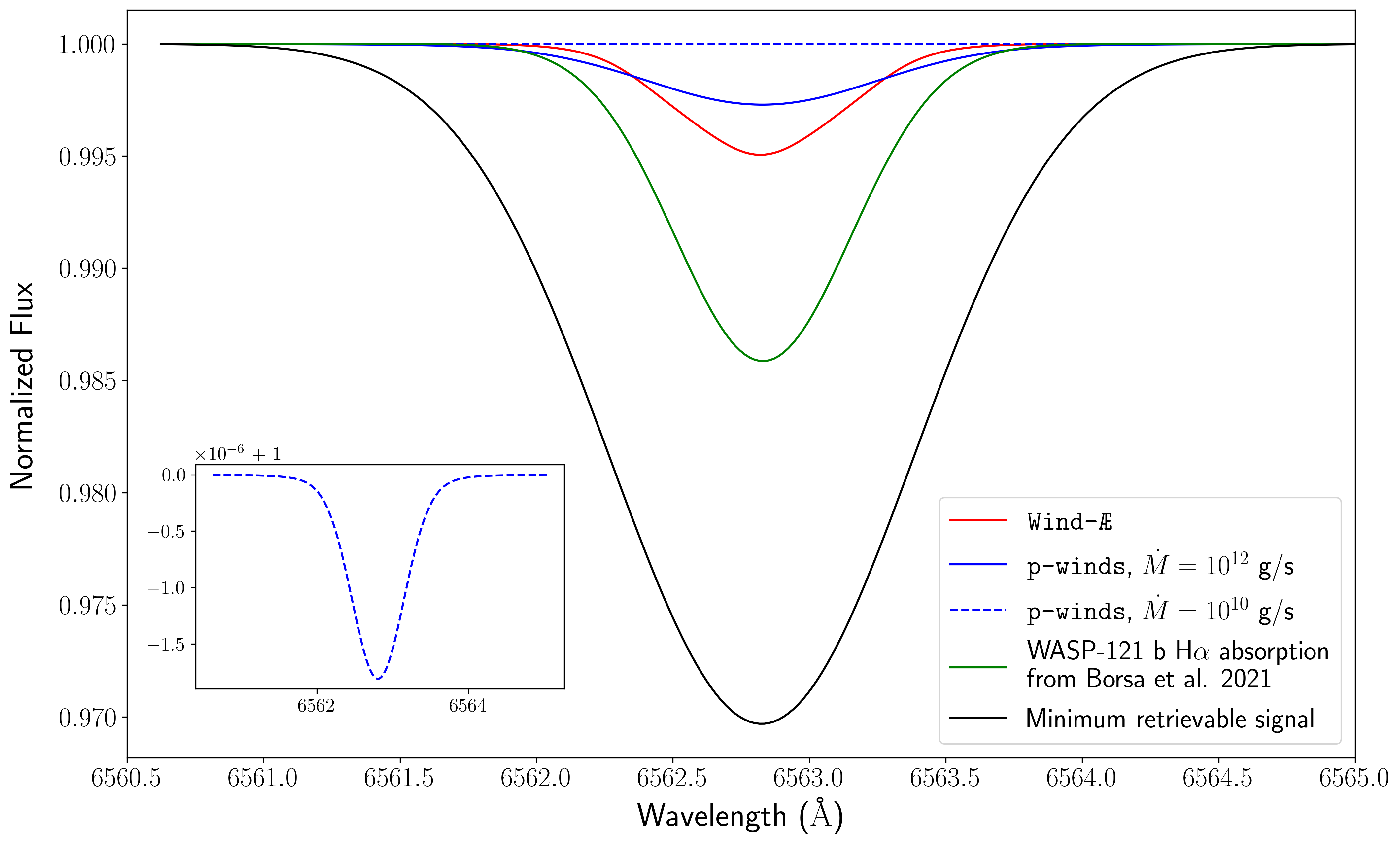} \\
    \caption{H$\alpha$ absorption feature calculated by \texttt{p-winds} for the weak ($10^{10}$ g s$^{-1}$, dashed blue, see inset) and moderate ($10^{12}$ g s$^{-1}$, solid blue) mass-loss rates as well as \texttt{Wind-\AE} (red) compared to the minimum retrievable signal (black, arbitrary width) from the Night 1 data set. We also include the best-fit model of WASP-121~b's empirical H$\alpha$ absorption (green) from \citet{Borsa2021} for comparison with observations of a similar system.}
    \label{fig:Halpha_pwind_signal}

    \begin{tabular}{ll}
        \hline
        \hline
        Injected H$\alpha$ model & SNR \\ 
        \hline
        \texttt{p-winds}, $\dot{M} = 10^{12}$ g s$^{-1}$ & 0.369 \\
        \texttt{Wind-\AE} & 0.676 \\
        WASP-121~b empirical H$\alpha$ absorption from \citet{Borsa2021} & 1.434 \\
        \hline
    \end{tabular}      
    \captionof{table}{SNRs from atmospheric escape model injection-recovery tests.}
    \label{tab:model_retrieved_SNR}
\end{figure}

From our injection-recovery analysis, we find that, under the assumptions of the \texttt{p-winds} model, we are unable to recover a $>$3$\sigma$ H$\alpha$ signal for either mass-loss case; we adopt the same SNR metric as in \S \ref{sec:Balmer_retrieval}. Figure \ref{fig:Halpha_pwind_signal} shows the H$\alpha$ absorption feature (the stronger of the two Balmer features we model) calculated by \texttt{p-winds} for the two mass loss cases according to the procedure described in \S \ref{sec:p-windsMethods}. The absorption signature for either mass loss case is weaker than the minimum amplitude signal with a comparable width (for this we adopt the signal corresponding to the purple star in Figure \ref{fig:Balmer_radialExtent}) that is retrievable at the 3$\sigma$ level given our data quality. As expected, we find that the modelled H$\beta$ feature is weaker than the H$\alpha$ feature and thus would make a detection of H$\beta$ more challenging than a detection of H$\alpha$. The H$\beta$/H$\alpha$ line depth ratio is 0.153 for the $\dot{M} = 10^{12}$ g s$^{-1}$ model and 0.138 for the $\dot{M} = 10^{10}$ g s$^{-1}$ model. Consequently, neither of the mass-loss cases are observable in either H$\alpha$ or H$\beta$. This is also highlighted in Table \ref{tab:model_retrieved_SNR}, which reports an SNR of 0.369 for the moderate mass loss \texttt{p-winds} model.

Given the lack of observability of Balmer features inferred from this injection-recovery analysis, we cannot use our \texttt{p-winds} models to place a meaningful constraint on WASP-12~b's mass-loss rate.  We do not attempt to increase the mass loss rate further to find a 3$\sigma$ upper limit on our constraint because WASP-12~b's mass loss rate is expected to be around $\dot{M}_{\mathrm{pl}} = 10^{11.4}$ g s$^{-1}$ in an energy-limited framework \citep{Ehrenreich2011}. \citet{Ehrenreich2011} estimate mass loss rates of the known transiting planets at the time of publication (including WASP-12~b) to be between $10^6$ and $10^{13}$ g s$^{-1}$, so we do not find it meaningful to attempt to constrain a higher mass loss rate than what we consider as an upper limit.

Additionally, the models presented thus far have many limitations as a consequence of certain simplifications:
\begin{enumerate}[1.]
\item By employing \texttt{p-winds}, we assume the outflow can be represented as a 1-D, isothermal Parker wind. In reality, the outflow is 3D and likely has a spatially-varying temperature gradient that can span thousands of Kelvin over the transit volume that we probe. 
\item The \texttt{p-winds} code does not account for molecular hydrogen, although this may not be a severe drawback since the extreme temperatures in the escape regime suggest any molecular hydrogen should be largely thermally dissociated. 
\item We approximate the high energy instellation upon the planet using the solar spectrum ($T_{\mathrm{eff}} = 5777 \pm 10$ K, \citet{Smalley2005}), which has a notably lower effective temperature than WASP-12 ($T_{\mathrm{eff}} = 6360^{+130}_{-140}$ K, \citet{Collins2017}).
\item While they make up a very small fraction of the atmospheric composition, metals do play a critical role in cooling and regulating the temperature structure of the outflow. The only metals accounted for in the \texttt{p-winds} model are carbon and oxygen. Furthermore, we assume solar abundances of carbon and oxygen. 
\item One major criticism of our models is that they require extreme temperatures to support physically plausible mass loss rates (our weak and moderate cases). For example, the lowest isothermal temperature profile at which \texttt{p-winds} converges to a solution for the moderate mass-loss case is at a temperature of 13,000 K; for the weak mass-loss case, it is closer to 11,000 K. These temperatures are unexpectedly high by a factor of $\sim$2 compared to expectations from hydrodynamic escape codes (see the solid orange curve in Figure \ref{fig:modelComparison}b). Ionization at such implausibly high temperatures is dominated by thermal (collisional) ionization, which is not included in \texttt{p-winds}, rather than photoionization. Furthermore, a higher temperature increases the Lyman-$\alpha$ emissivity in the atmosphere and may artificially inflate the model atmosphere, both of which can result in a larger H$\alpha$ transit depth. In reality, the absorption signals may be even weaker in amplitude if the physical conditions permit a lower temperature. 
\end{enumerate}


\subsection{Model injection-recovery with an upcoming multispecies relaxation code (\texttt{Wind-\AE}) provide a tentative upper limit}
\label{sec:relaxed-ae}

We adopt another code under development to model WASP-12~b's atmospheric escape with a more realistic (e.g. non-isothermal, multispecies, multifrequency) and flexible (e.g. tunable system parameters, fast) treatment of the outflow structure. \texttt{Wind-\AE}  (Broome et al., submitted) is a fast 1D photoionization atmospheric escape code adapted from \cite{rmc2008}. It is a relaxation code with new multispecies and multifrequency capabilities that models atmospheric escape as a transonic Parker wind. We expect \texttt{Wind-\AE} to yield more realistic outflow models than \texttt{p-winds} because it does not assume an isothermal atmosphere and can incorporate more metal species which significantly shape the thermal profile.


Boundary conditions for WASP-12~b include $T$($R_{\mathrm{min}}$) = 1000 K, $\rho$($R_{\mathrm{min}}$) = $5.67\times10^{-10}$ g cm$^{-3}$, where $T$ is temperature and $\rho$ is mass density at the user-defined minimum radius $R_{\mathrm{min}}$ = 1.02 $R_{\mathrm{pl}}$. We use the following planetary parameters: $M_{\mathrm{pl}}$ = 1.47 $M_{\mathrm{jup}}$,
$R_{\mathrm{p}}$ = 1.94 $R_{\mathrm{jup}}$,
$M_{*}$ = 1.43 $M_\odot$, and
$a$ = 0.02 AU,
which are the planet mass, planet radius in the UV, stellar mass, and semimajor axis respectively \citep{Chakrabarty2019}. We adopt an XUV range (10-2000 eV) flux of $F_{H}$ = 6.063 $\times\ 10^{4}$ erg cm$^{-2}$ s$^{-1}$ for scaling the solar spectrum \citep{FISM2020} to approximate the high energy spectrum of WASP-12. Note that some of these parameters, e.g. planetary radius, differ slightly from other parameters used in the \texttt{p-winds} model, for which we exclusively use the planet parameters derived in \citet{Collins2017} (listed in Table \ref{tab:petit_planet_params}). However, these differences are not significant enough to drastically change the resulting outflow models.

We assume solar metallicity \citep{lodders2009} and adopt the following species for the composition of the atmosphere: H, He, C, N, O, Ne, Mg, Si, Ca. We first generate fully self-consistent models out to $\sim$2.5 R$_{\mathrm{p}}$, the Coriolis turning radius of the planet; beyond this point, the assumption of spherical symmetry is no longer valid because the outflow no longer travels perpendicular to the planet's surface, but rather is turned by the Coriolis force into a tail. The planetary outflow is also subject to additional physics that we do not model at larger radii, such as charge exchange with the stellar wind which will dominate the energy and ionization of the planetary wind (\texttt{p-winds} also does not incorporate these effects). With these caveats in mind, we attempt to integrate the \texttt{Wind-\AE} model profile out to 20 R$_\mathrm{p}$ to match the radial range of the previous \texttt{p-winds} models. To facilitate the integration, it was necessary to artificially inflate the sonic point column density boundary condition by a factor of 2, corresponding to a difference in the resulting model mass loss rate of 6.4\% and near-negligible structural differences in the wind profile compared to the fully self-consistent number density model extending out to 2.5 R$_{\mathrm{p}}$. Upon conducting radiative transfer to model the Balmer lines, the discrepancy between the self-consistent 2.5 R$_{\mathrm{p}}$ model and the more unphysical 20 R$_{\mathrm{p}}$ model was negligible relative to observational uncertainties, so for the purposes of our investigation, we adopt the fully self-consistent 2.5 R$_{\mathrm{p}}$ model.

\begin{figure*}
    \centering
    \begin{minipage}{0.51\linewidth}
        \centering
        \includegraphics[width=\textwidth]{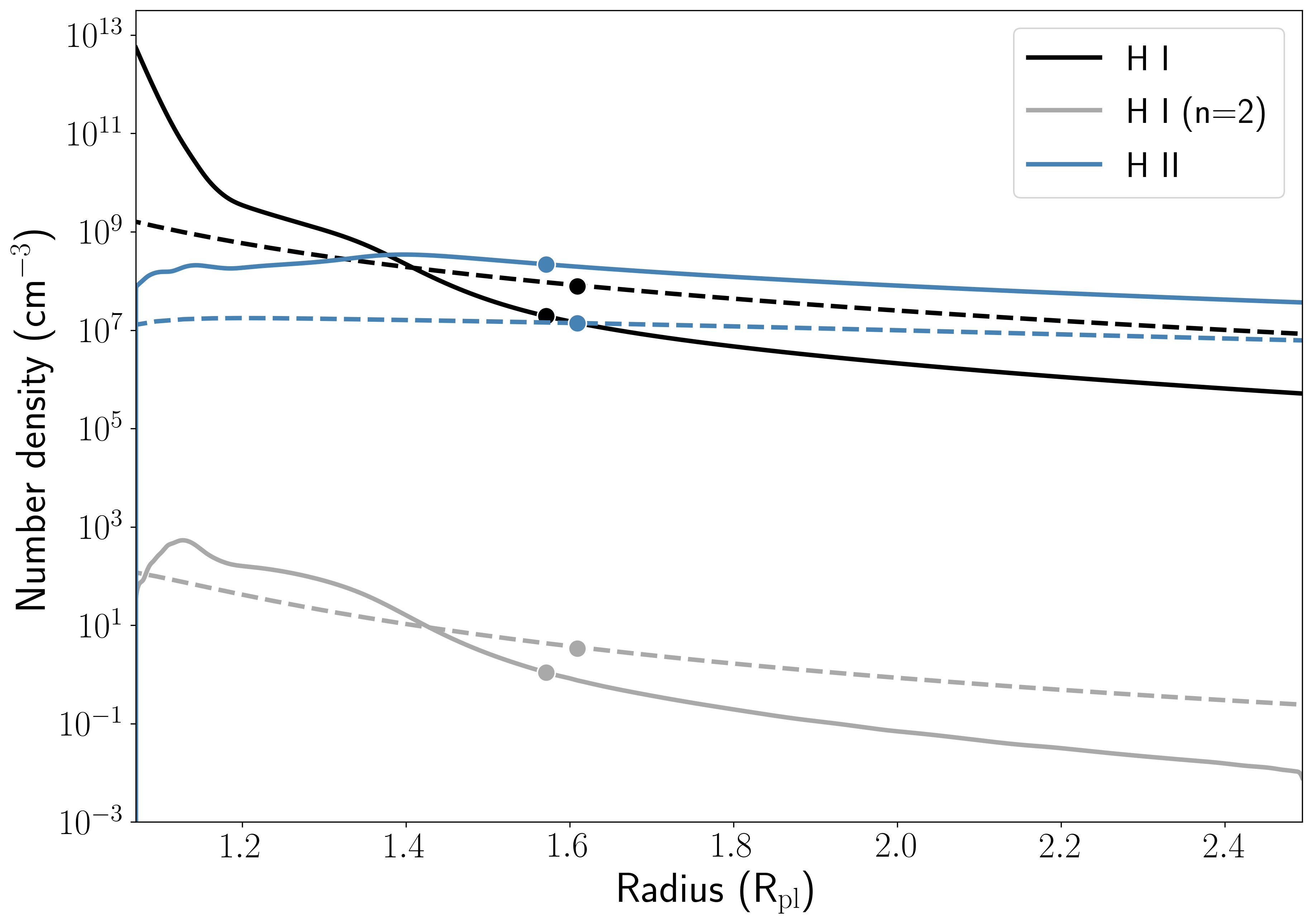} \\
        \label{fig:modelComparison_abundance}
        (a)
    \end{minipage}\hfill
    \begin{minipage}{0.48\linewidth}
        \centering
        \includegraphics[width=\textwidth]{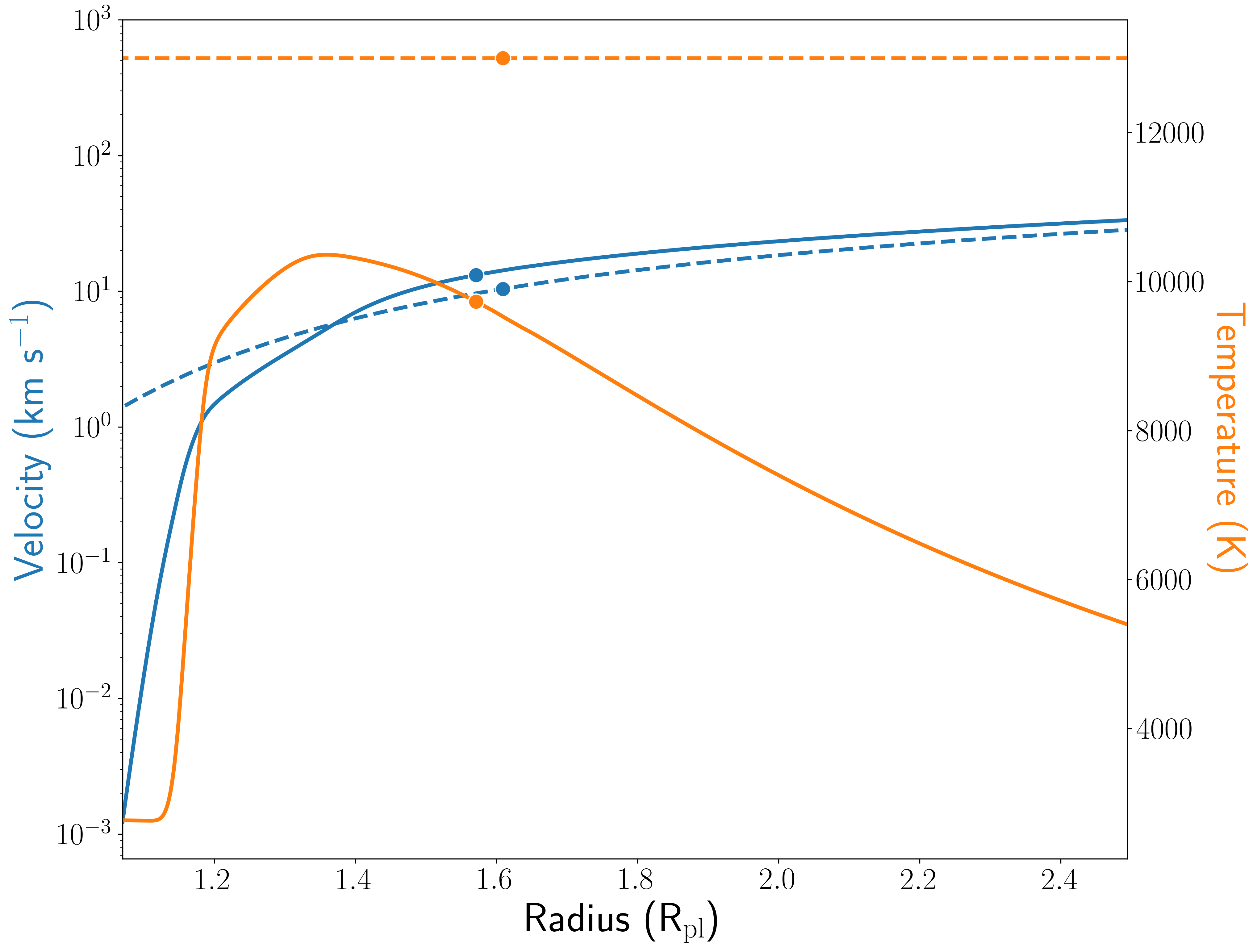} \\
        \label{fig:modelComparison_wind}
        (b)
    \end{minipage}
    \caption{Radial profiles of (a) hydrogen species number density and (b) outflow velocity and temperature for the \texttt{Wind-\AE} model in the solid curves and the $\dot{M} = 10^{12}$ g s$^{-1}$ \texttt{p-winds} model in the dashed curves.  The circular points mark the sonic point for the respective model of the curve they lie upon.} 
    \label{fig:modelComparison}
\end{figure*}

With \texttt{Wind-\AE}, the modelled planetary outflow has a mass-loss rate of $\dot{M} = 3.953 \times 10^{12}$ g s$^{-1}$ with the sonic point at 1.554 R$_{\mathrm{p}}$. In contrast to the \texttt{p-winds} models, the H$\beta$/H$\alpha$ line depth ratio from \texttt{Wind-\AE} is 0.267. A comparison of the hydrogen number density, wind velocity, and temperature profiles of WASP-12~b's outflow as modelled by \texttt{Wind-\AE} (solid curves) and \texttt{p-winds} (dashed curves) is provided in Figure \ref{fig:modelComparison}. Notably, \texttt{Wind-\AE} can incorporate more metal species and a more realistic, radially-varying temperature profile, lending more credibility to the model. In spite of these differences, it is interesting to note that the sonic point and velocity profiles at larger radii are not significantly discrepant between \texttt{Wind-\AE} ($R_{\mathrm{s}} = 1.571\ R_{\mathrm{pl}}$) and \texttt{p-winds} ($R_{\mathrm{s}} = 1.609\ R_{\mathrm{pl}}$ for $\dot{M} = 10^{12}$ g s$^{-1}$; $R_{\mathrm{s}} = 1.587\ R_{\mathrm{pl}}$ for $\dot{M} = 10^{10}$ g s$^{-1}$). Since these models incorporate tidal effects, the sonic point should be close to the L1 point, which is at 1.69 $R_{\mathrm{pl}}$ for WASP-12~b. 

The H-$\alpha$ feature resulting from the \texttt{Wind-\AE} model, given by the red curve in Figure \ref{fig:Halpha_pwind_signal}, is deeper than both the \texttt{p-winds} models, but does not fall below the minimum retrievable signal in black. When injected in our data from both nights, it corresponds to an SNR of 0.676 and is thus is not observable with our data quality.  Notably \citet{Czesla2024} place an upper limit on WASP-12~b's mass loss rate of $\lesssim 4 \times 10^{12}$ g s$^{-1}$, similar to the mass loss rated derived by \texttt{Wind-\AE}, from their non-detection of helium. However, we have shown that we do not have the data quality to validate this upper limit due to substantial photon noise in the optical with PEPSI/LBT. This may suggest that Balmer lines do not produce planetary spectral absorption signatures that are strong enough to be observed with current telescope facilities for a target as faint as the WASP-12 system ($V = 11.569$).

We also compare with the observed H$\alpha$ absorption signal of WASP-121~b from \citet{Borsa2021}. To first-order, the WASP-121 system is very similar to WASP-12 in terms of the spectral type and age (although poorly constrained, but very likely on the main sequence) of the host star as well as the equilibrium temperature and surface gravity of the planet. Consequently, we expect similar planetary outflow dynamics for the two systems. As indicated by the green curve in Figure \ref{fig:Halpha_pwind_signal} and the SNR of 1.434 reported in Table \ref{tab:model_retrieved_SNR}, if WASP-12~b possessed an H$\alpha$ signature comparable to that of WASP-121~b, the absorption feature would be insufficient to detect with both nights of our PEPSI observations stacked. Additionally, both the \texttt{p-winds} and \texttt{Wind-\AE} models fall short of the observed H$\alpha$ transit depth of WASP-121~b as seen in Figure \ref{fig:Halpha_pwind_signal}. This motivates further validation of both codes, since WASP-121~b's absorption features (including H$\alpha$) were successfully modelled with a robust, multispecies framework in \citet{Huang2023}. When compared to our \texttt{Wind-\AE} models of WASP-12~b, the models of WASP~121~b in \citet{Huang2023} display a shallower decrease in pressure below the temperature peak, resulting in a more inflated H$\alpha$ transit depth. Determining the root cause of this difference is beyond this scope of this observational work, and we recommend a future code comparison paper of atmospheric escape models commonly used in the field.

\section{Spectral Survey of Atomic and Molecular Features}
\label{sec:otherSpecies}

\subsection{Spectral survey of other atomic and molecular features via cross-correlation}
\label{sec:other_species}

We also search for other optical absorbers commonly observed in UHJ atmospheres, namely metal species such as Fe \RomanNumeral{1}/\RomanNumeral{2}, Ti \RomanNumeral{1}/\RomanNumeral{2}, Cr \RomanNumeral{1}/\RomanNumeral{2}, and potential agents of thermal inversions such as TiO and VO. We scour the wavelength range between 4900 \angstrom\ and 5400 \angstrom\ to avoid broad H$\beta$ absorption at lower wavelengths and tellurics at higher wavelengths. Many of the species we investigate have strong features in this wavelength range (see Figure \ref{fig:SNRmaps}). We generate template spectra assuming solar abundances over our specified wavelength range using \texttt{petitRADTRANS} \citep{Molliere2019} to  cross-correlate with our reduced transmission spectra, where the cross-correlation function (CCF) is defined according to Equation 13 in \citet{PaiAsnodkar2022}. We adopt a Guillot presssure-temperature profile \citep{Guillot2010} and set the reference pressure such that the resulting continuum of the template spectrum is close to the white-light transit depth. Table \ref{tab:petit_planet_params} lists the planetary parameters that we adopted in constructing all of the template spectra.

We generate individual templates for each species we considered so that any signals in the CCF can be solely attributed to that species. Thus, the abundances of the singular species of interest as well as hydrogen, and helium (since UHJ atmospheres are hydrogen-helium dominated) were inputs in the construction of any given template spectrum. The species that we focus on in this work have been validated as "detectable" by confirming that they have multiple lines in our wavelength range and yield a peak in the CCF when cross-correlated with observed stellar spectra. For the metal atomic species, we validate the template spectra against the observed spectrum of the host star WASP-12 from our PEPSI observations. For TiO and VO, we validate against archival spectra of GJ 793 (from HARPS-N) and LHS 2065 (from Keck HIRES\footnote{\url{https://koa.ipac.caltech.edu/cgi-bin/KOA/nph-KOAlogin}}) respectively, which are M-dwarfs known to display these features \citep{Gray2009}. Table \ref{tab:petit_species_params} specifies the mass mixing ratios adopted based on solar abundance \citep{Palme2014} and whether or not the species was identified as detectable according to our aforementioned metric. 

\begin{table*}
\centering
\caption{\texttt{petitRADTRANS} planetary parameter inputs.}
\begin{tabular}{llll}
\hline
\hline
Parameter & Units &  Value & Reference \\ 
\hline
Planet radius & $R_{\mathrm{jup}}$ & 1.9 & \citet{Collins2017} \\
Planet mass & $M_{\mathrm{jup}}$ & 1.47 & \citet{Collins2017} \\
Stellar radius & $R_{\odot}$ & 1.657 & \citet{Collins2017} \\
Planet surface gravity & m~s${^{-2}}$ & 10.09 & \citet{Collins2017}$^\dagger$ \\
Mean molecular weight &  & 2.33 & \\
Equilibrium temperature & K & 2580 & \citet{Collins2017} \\
Internal temperature & K & 100 & \\
Pressure range & bar & $10^{-10}$ -- $10^{2}$ & \\
Reference pressure & bar & 0.05 & \\
Infrared atmospheric opacity &  & 0.01 & \\
Ratio between optical and IR opacity &  & 0.4 & \\
\hline
\end{tabular}    
\label{tab:petit_planet_params}
\\ $^\dagger$Derived from reported planetary mass and radius.
\end{table*}

\begin{table}
\centering
\caption{\texttt{petitRADTRANS} atomic/molecular abundance inputs and corresponding detectability. Here detectability of a given species is marked with a green checkmark if cross-correlation between the template spectrum of the species over the observed wavelength range (PEPSI blue arm) and an appropriate stellar spectrum (i.e. a star known to host absorption features from the species of interest) yields a >3$\sigma$ signal; otherwise, the species is deemed undetectable over the wavelengths of observation and is marked with a red "x". The variable $x$ in the listed abundances for hydrogen and helium refers to the abundance of the species of interest used in the construction of a given template. All templates were constructed using Kurucz (\url{http://kurucz.harvard.edu/}) linelists except for TiO and VO, which came from the ExoMol linelists provided in \citet{McKemmish2016}.}
\begin{tabular}{ccc}
\hline
\hline
Species & $\log_{10}$ Abundance &  Detectability  \\ 
\hline
Al \RomanNumeral{1} & $-5.53$ & \textcolor{red}{\xmark} \\
B \RomanNumeral{1} & $-9.3$ & \textcolor{red}{\xmark} \\
Be \RomanNumeral{1} & $-10.62$ & \textcolor{red}{\xmark} \\
Ca \RomanNumeral{1} & $-5.67$ & \textcolor{ForestGreen}{\cmark} \\
Cr \RomanNumeral{1} & $-6.36$ & \textcolor{ForestGreen}{\cmark} \\
Fe \RomanNumeral{1} & $-4.52$ & \textcolor{ForestGreen}{\cmark} \\
Fe \RomanNumeral{2} & $-4.52$ & \textcolor{ForestGreen}{\cmark} \\
K \RomanNumeral{1} & $-6.89$ & \textcolor{red}{\xmark} \\
Li \RomanNumeral{1} & $-10.97$ & \textcolor{red}{\xmark} \\
Mg \RomanNumeral{1} & $-4.46$ & \textcolor{ForestGreen}{\cmark} \\
N \RomanNumeral{1} & $-4.14$ & \textcolor{red}{\xmark} \\
Na \RomanNumeral{1} & $-5.7$ & \textcolor{red}{\xmark} \\
Si \RomanNumeral{1} & $-4.48$ & \textcolor{red}{\xmark} \\
Ti \RomanNumeral{1} & $-7.1$ & \textcolor{ForestGreen}{\cmark} \\
Ti \RomanNumeral{2} & $-7.1$ & \textcolor{ForestGreen}{\cmark} \\
V \RomanNumeral{1} & $-8.0$ & \textcolor{red}{\xmark} \\
V \RomanNumeral{2} & $-8.0$ & \textcolor{red}{\xmark} \\
Y \RomanNumeral{1} & $-9.79$ & \textcolor{red}{\xmark} \\
TiO & $-7.1$ & \textcolor{ForestGreen}{\cmark} \\
VO & $-8.0$ & \textcolor{red}{\xmark} \\
\hline
H & $\log_{10}{(0.748 \times (1-x))}$ & \\
He & $\log_{10}{(0.250 \times (1-x))}$ &\\
\hline
\end{tabular}
\label{tab:petit_species_params}
\end{table}

After the template spectra are generated, they are cross-correlated with the transmission spectra. We conduct SYSREM on the resulting CCF maps, which is more effective than when applied to the reduced transmission spectra. We confirm this by injecting a template spectrum, e.g. Fe \RomanNumeral{1}, in the in-transit data (before applying any SYSREM) with both nights combined. We scale the Fe \RomanNumeral{1} template such that it yields a 5$\sigma$ signal when SYSREM (1 systematic, 100 iterations) is performed on the transmission spectra before cross-correlation. We find that we are unable to recover this signal if SYSREM is not performed altogether (this is true of all species we tested), likely due to imperfect normalization across observations, whereas we recover an improved 6.83$\sigma$ signal when SYSREM is performed after cross-correlation. This is because cross-correlation stacks the signal in the transmission spectra that match a given template spectrum. Thus, it is less sensitive to the stellar line core noise artifacts unless the species of interest has strong stellar absorption lines. Therefore, the streaks are not present in most of the CCF maps. Moreover, SYSREM is more effective at identifying systematics since these maps are smoother than the transmission spectra maps. When we apply SYSREM both before and after cross-correlation, we find the significance of the detection is only improved to 5.18$\sigma$, possibly due to the signal being washed out by too many applications of SYSREM. Likewise, increasing the number of systematics to more than 1 generally does not yield an increase in detection significance. Thus, we adopt one systematic when applying SYSREM after cross-correlation for all species considered except Fe \RomanNumeral{1}/\RomanNumeral{2}, and Mg \RomanNumeral{1}. We use two systematics for these species since their stellar absorption lines are stronger and thus noisier, therefore requiring more SYSREM systematics to minimize correlated noise across observations. 

\par As per standard practice \citep[e.g.][]{Nugroho2017, Kesseli2022}, we construct signal-to-noise ratio (SNR) maps for a grid of planetary orbital velocities ($K_\mathrm{p}$) and net velocity offset ($v_{\mathrm{sys}}$) values by stacking the in-transit CCFs in the planet's rest-frame to get the signal. We define the noise to be the standard deviation of the CCF at velocities beyond 1.5$v_\mathrm{rot}$ away from 0~km~s$^{-1}$, where $v_\mathrm{rot}$ is the rotational velocity of the planet assuming it is tidally locked ($9.05$~km~s$^{-1}$). With these definitions, we compute the SNR for a grid of ($K_\mathrm{p}$, $v_{\mathrm{sys}}$) pairs. A single peak with SNR > 5 in this $K_\mathrm{p}$-$v_{\mathrm{sys}}$ space qualifies as a detection of the corresponding species.

\subsection{WASP-12~b's atmosphere does not display optical absorption}

\begin{figure*}
\centering
\includegraphics[width=1\textwidth]{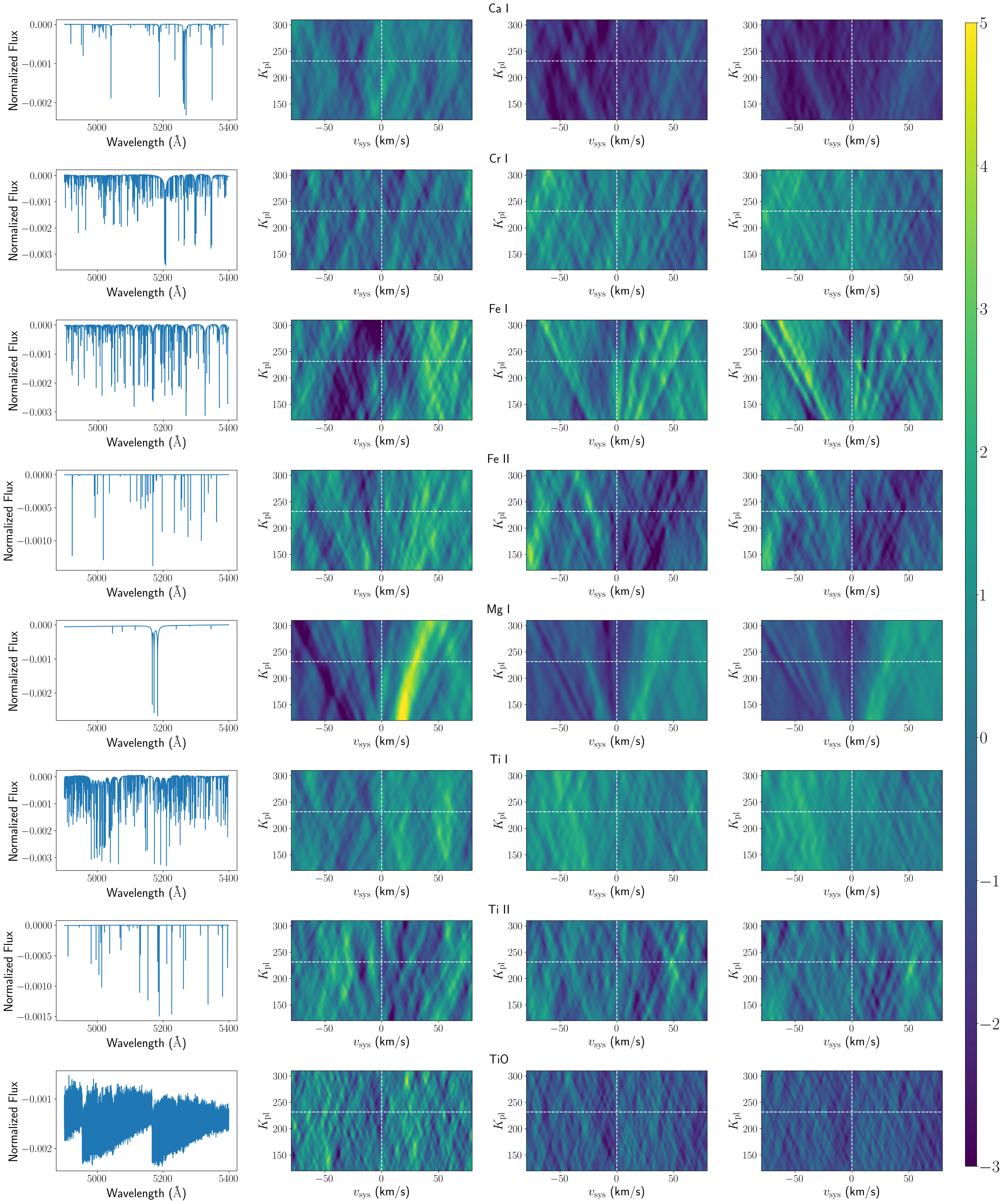}
\caption{(Columns from left to right) First column: template spectra for a given detectable atomic or molecular species in our wavelength range. Second column column: SNR maps of the corresponding species for Night 1. The dotted white lines correspond to the expected parameters of the system in the absence of velocity shifts from atmospheric dynamics. The colorbar represents SNR from stacking the cross-correlated spectra in a planetary rest-frame of the corresponding $v_{\mathrm{sys}}$ and $K_{\mathrm{pl}}$ combination represented along the x and y axes respectively. Third column: same as second column, but for Night 2. Fourth column: same as second column, but for both nights stacked. The panels in the last three columns displaying our search for absorption do not show detections from the planetary atmosphere for any of the investigated tracers.}
\label{fig:SNRmaps}
\end{figure*}

\begin{figure}
\centering
\includegraphics[width=0.5\textwidth]{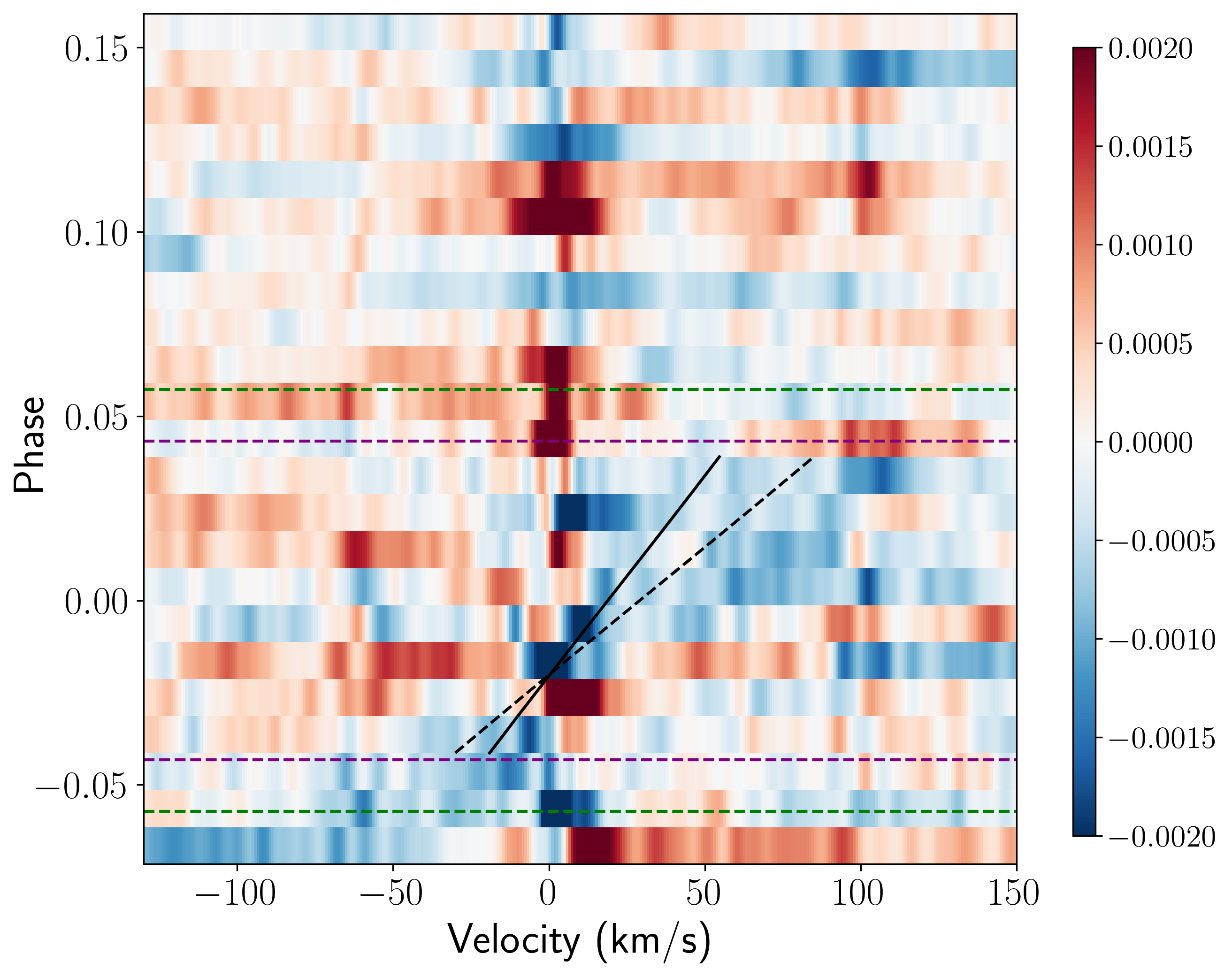} \\
\label{fig:MgI_Night1}
\caption{Cross-correlation with Mg \RomanNumeral{1} template for Night 1. As before, the green horizontal dashed lines indicate the phases of 1st and 4th contact while the purple horizontal dashed lines indicate the phases of 2nd and 3rd contact. The black solid line indicates the center of a purported absorption signature with $K_\mathrm{p}$ and $v_{\mathrm{sys}}$ corresponding to the peak of the Mg I SNR map for Night 1. The black dashed line corresponds to the absorption track with the highest SNR value of $v_{\mathrm{sys}}$ for a $K_\mathrm{p}$ that matches the projected orbital velocity of the planet.  The vertical stationary feature near a velocity of 0~km $\mathrm{s}^{-1}$ corresponds to the residual signal from the stellar line core.}
\label{fig:MgI}
\end{figure}


Our search for absorption in the blue arm PEPSI data from various atomic and metal species using both nights of observation does not yield any detections; see Figure \ref{fig:SNRmaps}. The Mg \RomanNumeral{1} map for the first night shows a spurious signal near $K_{\mathrm{pl}}$ = 150 km~$\mathrm{s}^{-1}$ and $v_{\mathrm{sys}}$ = 20 km~$\mathrm{s}^{-1}$. Figure \ref{fig:MgI} shows the Mg \RomanNumeral{1} cross-correlation maps for Night 1 with the purported absorption track corresponding to the peak of the Mg I SNR map traced out with a black solid line. The signal in the Mg \RomanNumeral{1} SNR map for Night 1 could be purely coincidental or an alias with the noisy stellar line cores since the Mg \RomanNumeral{1} spectrum in this wavelength range is dominated by just three lines in the triplet between 5167-5183 \AA. Furthermore, the signal's corresponding value of $K_{\mathrm{pl}}$ is nearly 100 km~s$^{-1}$ less than expected for this planet (231.5 km~s$^{-1}$) and thus this signal cannot be attributed the planet's atmospheric absorption.


The Mg \RomanNumeral{1} b triplet at 5167, 5172, and 5183 \angstrom\ is of particular note as a potential tracer of atmospheric escape \citep{Cauley2019}. We do not observe such absorption of planetary origins from either an inspection of the constructed transmission spectra nor cross-correlation. Other lines of potential interest to constraining mass-loss include Ca \RomanNumeral{1} $\lambda$4227 \angstrom, Mg \RomanNumeral{1} $\lambda$4571 \angstrom\, Na D doublet $\lambda$5890, 5896 \angstrom, and Ca \RomanNumeral{1} $\lambda$6122; however, these are outside our wavelength range with the PEPSI cross-dispersers we have chosen for these observations, but can be achieved with others \citep{Keles2024}. In addition to the absorbers presented, we also tested the following atomic and molecular species that are immediately available from the \texttt{petitRADTRANS} high-resolution opacity database: Al \RomanNumeral{1}, B \RomanNumeral{1}, Be \RomanNumeral{1}, K \RomanNumeral{1}, Li \RomanNumeral{1}, N \RomanNumeral{1}, Na \RomanNumeral{1}, Si \RomanNumeral{1}, V \RomanNumeral{1}, V \RomanNumeral{2}, Y \RomanNumeral{1}, and VO. These species were deemed undetectable through cross-correlation with an appropriate empirical stellar spectrum (see Table \ref{tab:petit_species_params}). This is because they lack sufficient absorption signal in our wavelength range, with only one or a few lines that are prone to aliasing when cross-correlated.

To compare our non-detections with the literature, we note that \citet{Burton2015}, \citet{Jensen2018} and \citet{Czesla2024} are the only other works we could find that conduct high-resolution transmission spectroscopy of WASP-12~b in the optical. In addition to their H$\alpha$ detection, \citet{Jensen2018} observes planetary Na \RomanNumeral{1} absorption via the Na D doublet; \citet{Burton2015} also claims a tentative detection of Na D doublet absorption in the atmosphere of WASP-12~b using defocused transmission spectroscopy. Previous works applied ultraviolet spectroscopy to infer metals in WASP-12~b's exosphere such as Na \RomanNumeral{1},  Sn \RomanNumeral{1}, Mn \RomanNumeral{1}/\RomanNumeral{2}, Yb \RomanNumeral{2}, Sc \RomanNumeral{2}, Al \RomanNumeral{2}, V \RomanNumeral{2}, Mg \RomanNumeral{2} \citep{Fossati2010, Haswell2012}; we are not sensitive to any of these in our wavelength range. They also suspect the presence of other metals that we have searched for, namely Mg \RomanNumeral{1} and Fe \RomanNumeral{1}. \citet{Fossati2010} do not provide abundance estimates, so we are unable to consistently constrain the expected observability of their atomic line measurements in our optical wavelength regime. Nonetheless, our non-detections at high-resolution present a challenge to reconcile with UV constraints on WASP-12~b's exospheric composition. Fully understanding this discrepancy will require extensive modelling of the outflow across UV and optical wavelengths. Our non-detections in the optical placed in the context of the literature presents an opportunity to consolidate a holistic, multiwavelength understanding of this system. 

We also note that our observations are taken with exposure times of 900 s per spectrum to beat down photon noise. Consequently, the planet changes in radial velocity by $\sim$14 km s$^{-1}$ ($\sim 6 \times$ greater than the width of PEPSI resolution element, 2.3 km s$^{-1}$) over the course of a single observation. Thus this orbital motion smears out any potential planetary absorption feature, both Balmer lines and atomic metal lines. This motion blur may drive the absence of atmospheric absorption signatures in our optical data.



\subsection{Comparison with other UHJs}
The absence of optical absorption features also presents a challenge for comparative planetology of UHJ atmospheres. Atomic metal line absorption from Fe \RomanNumeral{1}/\RomanNumeral{2}, Cr \RomanNumeral{1}/\RomanNumeral{2}, Na \RomanNumeral{1}, and a plethora of other species has become a characteristic finding in UHJ atmospheres like WASP-76~b, KELT-9 b, KELT-20 b, and more \citep{Hoeijmakers2019, Casasayas-Barris2019}. At the same time, many UHJs lack such detections despite attempts to search for them with transmission spectroscopy, such as HAT-P-57~b, KELT-7~b, KELT-17~b, KELT-21~b, MASCARA-1b, and others \citep{Strangret2022}. The case of KELT-7~b is particularly noteworthy since this planet is around a star with an effective temperature comparable to WASP-12's. However, KELT-7~b also has a smaller scale height that is less amenable to observation due to its significantly higher surface gravity and lower equilibrium temperature. WASP-12~b is also similar to WASP-76 b and WASP-121 b insofar as their equilibrium temperatures and host star spectral types, yet the latter pair display numerous optical atomic and molecular features in transmission \citep{Kesseli2022, Sanchez-Lopez2022b, Seidel2019, Pelletier2022, Cabot2020b, Ben-Yami2020, Gibson2020, Hoeijmakers2020}. WASP-76~b has a larger scale height, so this may explain why its atmospheric absorption features are observable when WASP-12~b's are not. However, WASP-121~b may have a scale height smaller than WASP-12~b's depending on the literature value adopted for its equilibrium temperature.

\section{Conclusions}
\label{sec:conclusions}

We have presented a search for atmospheric absorption in PEPSI-LBT high-resolution optical transmission spectra of WASP-12~b and report no evidence of planetary absorption features. Our lack of an H$\alpha$ detection is in direct contradiction with a previous observation at medium resolution in \cite{Jensen2018}, but we show that this detection will require exotic phenomena if it is planetary in origin. We conduct injection-recovery tests to constrain the radial extent and escape rate of WASP-12~b's hydrogen envelope. This analysis suggests that we do not have the sensitivity to determine if the hydrogen envelope is confined within the planet's Roche lobe. We explore 1-D models of a planetary outflow using the \texttt{p-winds} and \texttt{Wind-\AE} codes. The modelled Balmer line absoprtion features are much smaller in amplitude than the photon noise of our observations as well as the H$\alpha$ signal purported in \cite{Jensen2018}.

From a theoretical standpoint, WASP-12~b possesses qualities that are favorable for observing atmospheric escape. \citet{Koskinen2022} show that even only considering Roche lobe overflow, the system parameters of WASP-12~b suggest that it should have one of the highest mass-loss rates amongst the known planets; this is supported by hydrodynamic outflow inferred from the observed in-transit excess UV absorption. However, the WASP-12 system is fainter than many canonical UHJs observed at high resolution. Our non-detection of Balmer line absorption with the LBT, a telescope at limit of current facilities with the highest light-collecting area, poses a challenge for further observations of this target at high resolution in the optical. Further investigation of this target with PEPSI will require stacking more observations or observing at lower spectral resolution.


We also search for other optical absorbers including several metal species and TiO. We find no absorption features despite previous inferences of metals in WASP-12~b's exosphere from excess UV absorption in transit. Reconciling WASP-12 b's seemingly barren optical atmosphere in the context of the ever-growing repository of high-resolution detections in UHJ atmospheres will require deeper insight from sophisticated modelling and further observation to tighten empirical constraints.

\section*{Acknowledgements}

This work is supported by the National Science Foundation under Grant No. 2143400. A.P.A would like to thank the David G. Price Fellowship in Astronomical Instrumentation and the NASA Space Technology Graduate Research Opportunity (NASA Grant 80NSSC22K1197) for funding her over the course of writing this work. MCJ is supported by NASA Grant 80NSSC23K0692. A.P.A is also extremely grateful for Professor Adam Leroy's Interstellar Medium graduate course lecture notes, which were foundational reference materials. The authors also thank Jack Neustadt and Romy Rodr\'{i}guez Mart\'{i}nez for providing their insight on the selection of stellar spectral types with the necessary spectral features for validating our TiO/VO linelists. 


This work is based on observations made with the Large Binocular Telescope. The LBT is an international collaboration among institutions in the United States, Italy and Germany. LBT Corporation partners are: The University of Arizona on behalf of the Arizona Board of Regents; Istituto Nazionale di Astrofisica, Italy; LBT Beteiligungsgesellschaft, Germany, representing the Max-Planck Society, The Leibniz Institute for Astrophysics Potsdam, and Heidelberg University; The Ohio State University, representing OSU, University of Notre Dame, University of Minnesota and University of Virginia. We thank queue observers Dr. Jack Neustadt, Dr. Patrick Vallely, and Professor Rick Pogge as well as telescope operator Steve Allanson and special assistants Dr. Olga Kuhn and Alexander Becker for assisting with the collection of the PEPSI data presented in this work.

This work is also based on observations obtained with the Hobby-Eberly Telescope (HET), which is a joint project of the University of Texas at Austin, the Pennsylvania State University, Ludwig-Maximillians-Universitaet Muenchen, and Georg-August Universitaet Goettingen. The HET is named in honor of its principal benefactors, William P. Hobby and Robert E. Eberly. We acknowledge the Texas Advanced Computing Center (TACC) at The University of Texas at Austin for providing high performance computing, visualization, and storage resources that have contributed to the results reported within this paper.

We thank the creators of open-source Python \citep{python} and IDL \citep{idl} software that power the analysis conducted in this work, including: \texttt{jupyter} \citep{jupyter}, \texttt{numpy} \citep{numpy}, \texttt{scipy} \citep{scipy}, \texttt{astropy} \citep{astropy:2013, astropy:2018, astropy:2022}, \texttt{matplotlib} \citep{matplotlib}, \texttt{pandas} \citep{pandas:paper}, \texttt{emcee} \citep{Foreman-Mackey2013}, \texttt{corner} \citep{corner}, \texttt{p-winds} \citep{DosSantos2022}, and \texttt{SME} \citep{Valenti1996, Valenti2012}.

\section*{Data Availability}

LBT/PEPSI data analyzed in this work will be provided upon request to the corresponding author. HET observations of the WASP-12 system belong to and were provided by contributing author Professor Adam Jensen.



\bibliographystyle{mnras}
\bibliography{main} 

\bsp	
\label{lastpage}
\end{CJK*}
\end{document}